# Low-power/high-gain flexible complementary circuits based on printed organic electrochemical transistors

Chi-Yuan Yang[1], Deyu Tu[1], Tero-Petri Ruoko[1], Jennifer Y. Gerasimov[1], Han-Yan Wu[1], P. C. Harikesh[1], Renee Kroon[1,2], Christian Müller[3], Magnus Berggren[1,2,4], Simone Fabiano[1,2,4]∗

[1]Laboratory of Organic Electronics, Department of Science and Technology, Linköping University, SE-601 74 Norrköping, Sweden.

[2]Wallenberg Wood Science Center, Linköping University, SE-601 74 Norrköping, Sweden.

[3]Wallenberg Wood Science Center, Department of Chemistry and Chemical Engineering, Chalmers University of Technology, SE-412 06 Göteborg, Sweden.

[4]n-Ink AB, Teknikringen 7, SE-583 30 Linköping, Sweden.

Correspondence should be addressed to: simone.fabiano@liu.se

**Abstract**

The ability to accurately extract low-amplitude voltage signals is crucial in several fields, ranging from single-use diagnostics and medical technology to robotics and the Internet of Things. The organic electrochemical transistor, which features large transconductance values at low operation voltages, is ideal for monitoring small signals. Its large transconductance translates small gate voltage variations into significant changes in the drain current. However, a current-to-voltage conversion is further needed to allow proper data acquisition and signal processing. Low power consumption, high amplification, and manufacturability on flexible and low-cost carriers are also crucial and highly anticipated for targeted applications. Here, we report low-power and high-gain flexible circuits based on printed complementary organic electrochemical transistors (OECTs). We leverage the low threshold voltage of both p-type and n-type enhancement-mode OECTs to develop complementary voltage amplifiers that can sense voltages as low as 100 µV, with gains of 30.4 dB and at a power consumption < 2.7 µW (single-stage amplifier). At the optimal operating conditions, the voltage gain normalized to power consumption reaches 169 dB/µW, which is >50 times larger than state-of-the-art OECT-based amplifiers. In a two-stage configuration, the complementary voltage amplifiers reach a DC voltage gain of 193 V/V, which is the highest among emerging CMOS-like technologies operating at supply voltages below 1 volt. Our findings demonstrate that flexible complementary circuits based on printed OECTs define a power-efficient platform for sensing and amplifying low-amplitude voltage signals in several emerging beyond-silicon applications.

Complementary metal-oxide-semiconductor (CMOS) field-effect transistors, made from silicon (Si), have been the workhorse of the integrated circuit (IC) industry since the 1980s, in part due to their low power consumption. Dennard scaling[1], the CMOS scaling law, states that the supply voltage for each new CMOS generation is reduced by 30%, and the power consumption subsequentially reduces by 50%. After decades of development, the latest 7-nm-node CMOS process reaches a supply voltage of 0.75 V[2]. Today, the Si-CMOS technology is heavily explored in Internet of Things (IoT) applications, serving as low-power outposts that record physical sensor parameters (e.g., motion, light, temperature), communicate over long distances, and harvest and store energy for its operation[3]. Expanding IoT modules with flexible, soft, or large-area chemical sensors and actuators only possible off-Si, enables a circuit technology that can amplify and route signals, facilitating signal compatibility and low-cost integration between Si-technology and embedded devices. Further, for many IoT and bioelectronic applications (e.g., (bio-)chemical sensors and neuronal interfacing), the on-site technology is preferably realized without Si-chips to enable many different form factors, proximity, elasticity, and signal transduction, tailor-made for the actual chemical/biological environment. Also in this case, a low-power/voltage, high-performing, and flexible circuit technology operating at the site of stimulation or sensing is needed to record and transfer signals at high signal-to-noise performance.

Unlike field-effect transistors, organic electrochemical transistors (OECTs) typically operate at less than 1 V and consist of a conducting polymer channel and a gate connected by a common electrolyte[4]. When a voltage is applied to the gate, ions from the electrolyte enter the bulk of the channel material to compensate for the injected charge carriers in the oxidized or reduced organic semiconductor, thus modulating the channel conductance. Since OECTs operate at low voltages and exhibit transconductance values that are orders of magnitude higher than their (organic) field-effect transistor counterparts, they are ideally suited to sense and amplify low-amplitude voltage signals[5]. To date, OECTs have been used to construct digital circuits[6–8], sensors of biological, physical, and chemical signals[9–11], and neuromorphic computing devices[12,13].

While the large OECT transconductance can translate a small voltage into a sizable current, post-signal processing often necessitates further current-to-voltage conversion. To address this need, a handful of high-gain OECT-based voltage amplifiers have been developed[14–20]. For instance, Braendlein et al. reported a voltage amplifier that consists of a depletion-mode OECT based on the hole-transporting (p-type) polymer poly(3,4-ethylenedioxythiophene):poly(styrenesulfonate) (PEDOT:PSS) and a resistor load to record ECG signals[16], while Romele et al. demonstrated voltage gains over 100 using crystallized PEDOT:PSS-based OECTs[18]. However, amplifiers based on depletion-mode OECTs almost universally suffer from high power consumption, in some instances reaching values as high as 1350 µW[16]. While efforts have been made to reduce the power consumption of OECT-based amplifiers down to 20 µW, this is usually done at the expense of the voltage gain (~9.3 V/V), as shown recently for enhancement-mode OECTs operated in the subthreshold regime[17].

In this work, we report a different approach to reduce power consumption, which takes advantage of CMOS technology. Si-based electronics have already shown significant improvements in power efficiency afforded by CMOS circuits compared to unipolar technologies, wherein only one type of transistor is used[21]. In order to develop CMOS-like OECT technology, both p-type and n-type OECTs working in enhancement mode are required[20]. Printing technologies, like ink-jet printing and screen-printing, are compatible with organic electronic materials and devices to enable flexibility, conformability, large-scale integration, and cost efficiency[8,22]. However, despite prior reports of large-scale circuits based on all-printed unipolar OECTs[6], printed complementary OECT amplifiers have not yet been reported.

Here, we report a printed, flexible single-stage voltage amplifier based on a pair of complementary OECTs, which operates at low voltages (0.3-0.7 V) with less than 2.7 µW power consumption. Both p-type and n-type OECTs operate in the accumulation mode, with maximum transconductance of ~0.20 mS and threshold voltages of less than |0.25| V, enabling low voltage operation. Electrochemical bulk doping of the channel materials is confirmed by in-situ spectroelectrochemistry measurements. The driving strength and the operating voltage of both p-type and n-type OECTs are well balanced and enable the development of single-stage complementary inverters having voltage gains of up to 26 V/V, static power consumption as low as 12 nW, and excellent noise margin (89%). With a DC offset at the input, the inverter operates as a pull-push amplifier which is able to sense AC voltage signals as low as 100 µV with a gain of 30.4 dB. This ability to detect small voltage signals with very low power consumption yields a power-normalized gain of up to 169 dB/µW, which is over 50 times greater than state-of-the-art amplifiers based on OECT technologies. The DC voltage gain can be further increased by cascading double inverters into a two-stage amplifier that reaches 193 V/V, which is the highest gain among emerging CMOS-like thin-films technologies operated at low voltages (sub-1-volt) to date. Our voltage amplifier based on printed complementary OECTs offers a power-efficient solution for autonomous, conformable, wearable, and portable sensors. In addition, the low operation voltage (0.3 V) offers the possibility of being (self-)powered by light, heat, wireless power, triboelectricity, and other power sources that can only provide low voltage and/or limited power supply, thus opening the way to battery-free wearable electronics.

**Printed complementary organic electrochemical transistors**
Complementary electronic circuitries require the development of both p-type and n-type enhancement-mode OECTs. Here, we use polythiophene functionalized with tetraethylene glycol side chains (P($g_4$2T-T)) and poly(benzimidazobenzophenanthroline) (BBL) as the p-type and n-type semiconducting polymers, respectively (Fig. 1a). While neither P($g_4$2T-T) nor BBL are soluble in water/alcohols, they can be dispersed in alcohol solvents in the form of nanoparticle inks, which are better suited for large-scale printing. The BBL nanoparticles are obtained by solvent exchange from BBL-methanesulfonic acid (MSA) solution to isopropanol (IPA) under rapid stirring, whereas the P($g_4$2T-T) nanoparticles can be obtained by solvent exchange from its chloroform solution to IPA. The nanoparticle sizes of BBL and P($g_4$2T-T) are 28 nm and 21 nm, respectively, as measured by dynamic light scattering (DLS, Supplementary Figure 1). The BBL and P($g_4$2T-T) nanoparticle dispersion inks in IPA are printable by both screen printing and inkjet printing. We chose a poly(sodium-4-styrene sulfonate) (PSSNa, Fig. 1a) based hydrogel as the printable sodium electrolyte for the n-channel OECT. *D*-sorbitol and glycerin are employed to enhance sodium conductivity and the stability of the PSSNa-based hydrogel. We chose polyquaternium-10 (PQ-10, a non-toxic quaternized hydroxyethyl cellulose chloride, Fig. 1a) based hydrogel as the printable chloride electrolyte for the p-channel OECT. Both PQ-10 and PSSNa hydrogels have similar or even higher ion conductivity than 0.1 M NaCl aqueous solution (Supplementary Figure 2). The all-printed OECT-based circuits are fabricated through the combination of screen printing and inject printing (Fig. 1b). First, carbon electrode and silver electrode layers are screen printed sequentially on a flexible A3 sized polyethylene terephthalate (PET) substrate. The electrochemically inert carbon electrodes are in direct contact with the polymer semiconductor, while the silver underlayer reduces the electrode resistance. An insulating layer is used to pattern the channel and gate regions.[6] The silver/silver chloride (Ag/AgCl, 100 nm thick) gate layer and polymer semiconductor layer (20 nm for P($g_4$2T-T) and 250 nm for BBL, to match their ON-state channel conductances) are sequentially deposited by spray-coating. Finally, the polymer hydrogel electrolytes (PQ-10 or PSSNa) are screen printed to finish the fabrication of the electrochemical circuits (Fig. 1c). For each OECT, the channel length/width (L/W) is 200 μm/2 mm, and the Ag/AgCl gate region is 2×2 mm$^2$. When implemented as OECT channels, both P($g_4$2T-T) and BBL show relatively low threshold voltages[23,24], ensuring that a low supply voltage is required for operation. The device structure of the printed complementary OECTs are depicted in Fig. 1b-d. Here, three layers, including a carbon layer for drain/source contact, a silver layer for interconnects, and an insulating layer to define the gate and channel windows, are sequentially screen-printed on a polyethylene terephthalate (PET) substrate (Supplementary Figure 3). Semiconductor layers (P($g_4$2T-T) and BBL) are then successively spray-coated through a shadow mask to form the transistor channels. A layer of Ag/AgCl is deposited at the silver gate electrodes and the electrolyte hydrogels (PQ-10 and PSSNa) are finally deposited to bridge the gate and the channel. More details of device fabrication can be found in the Methods section. The printed OECTs have a side-gate configuration, which is optimal for low-cost and large-scale manufacturing, as this outline requires fewer processing steps compared to top-gate configurations.

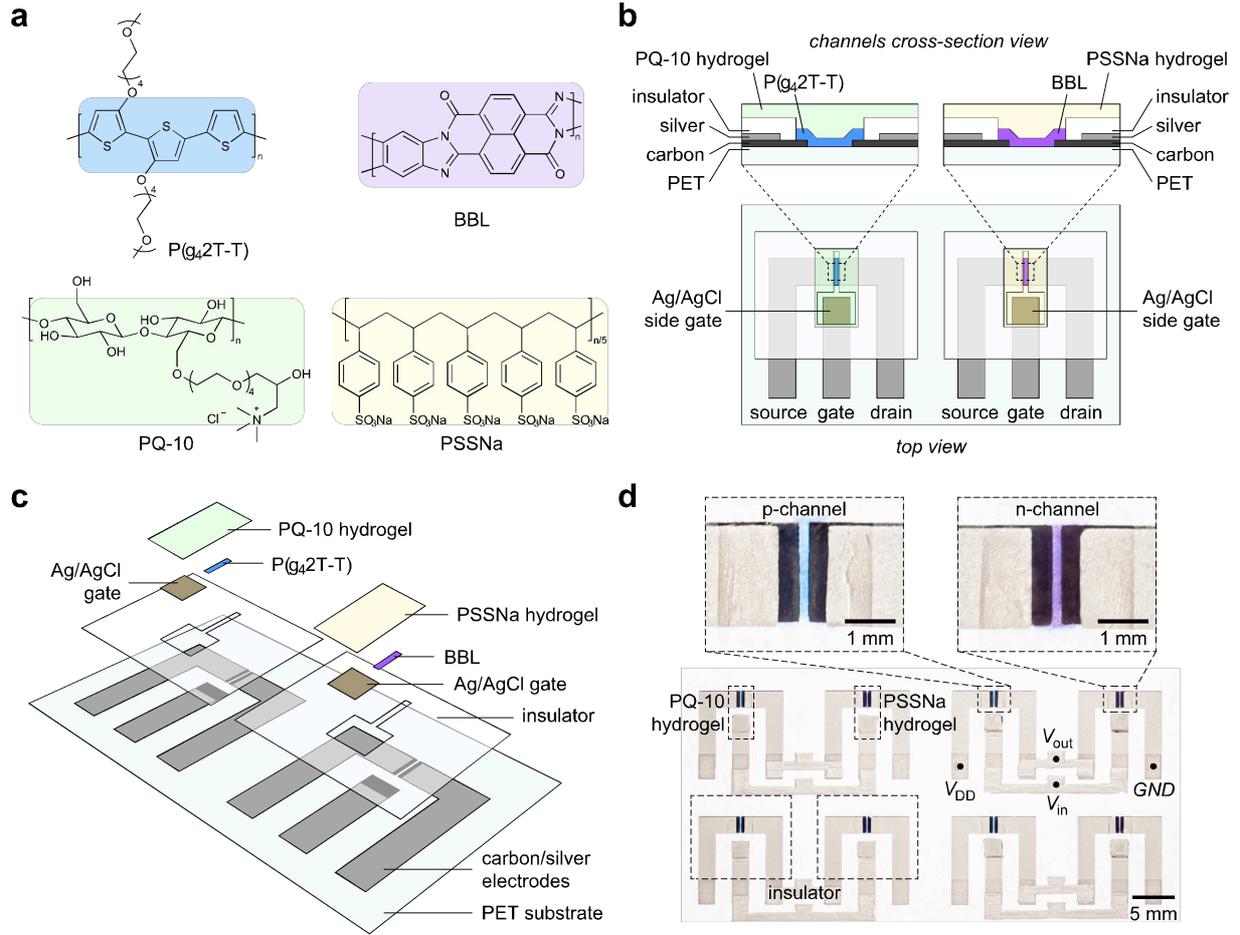

**Fig. 1 | Schematic of the complementary OECTs. a,** The molecular structures of p-type polymer P(g$_4$2T-T) and n-type polymer BBL and their corresponding electrolytes, PQ-10 and PSSNa. **b,** The illustrated structure of the complementary OECTs (cross-section view and top view). **c,** The breakdown of each layer in the fabrication process of the complementary OECTs. **d,** Photograph of the as-manufactured complementary OECTs on PET substrate.

### Electrical characteristics of printed p-/n-type OECTs

In a three-terminal (drain, source, and gate) OECT device (Fig. 1b), a bias voltage between gate and source ($V_{GS}$) drives ions from the electrolyte into the channel and results in the modulation of the conductivity between drain and source. For enhancement-mode OECTs, the drain-to-source current ($I_{DS}$) under a voltage bias ($V_{DS}$) can be approximately expressed as:[5]

$$I_{DS} = \mu C_V \frac{Wd}{L}\left[(V_{GS} - V_T)V_{DS} - \frac{V_{DS}^2}{2}\right] \quad (linear\ region) \quad (1)$$

$$I_{DS} = \mu C_V \frac{Wd}{2L}(V_{GS} - V_T)^2 \quad (saturation\ region) \quad (2)$$

where $\mu$ is the charge carrier mobility, $C_V$ is the volumetric capacitance, $W$, $L$, and $d$ are the channel width, length, and thickness, respectively, and $V_T$ is the threshold voltage. Unlike conventional field-effect transistors, the channel thickness $d$ contributes to the conductivity as the accumulation of charges occurs throughout the entire bulk of the semiconductor layer.

For printed P(g$_4$2T-T) OECTs, the application of a negative $V_{GS}$ oxidizes the semiconductor channel and enables the accumulation and transport of holes in the channel under a negative $V_{DS}$. Figs 2a-b show the typical output ($I_{DS}$-$V_{DS}$) and transfer ($I_{DS}$-$V_{GS}$) characteristics of P(g$_4$2T-T)-based OECTs reaching an ON/OFF ratio of ~10$^3$ at $V_{GS}$ < 1 V. In addition, no hysteresis in $I_{DS}$ is observed over the entire $V_{GS}$ range, which is beneficial for the stable operation of analog circuits (e.g., amplifiers). The transconductance $g_m = \partial I_{DS}/\partial V_{GS}$, a key figure of merit used to evaluate OECTs, is also shown in Fig. 2b. The maximal transconductance of P(g$_4$2T-T) OECTs is ~0.17 mS. From the transfer characteristics in the saturation region, a typical $V_T$ of -0.22 V is extracted (Supplementary Figure 4a). From Eq. (2) we can then extract the product of mobility and volumetric capacitance $\mu C_V$ = 22.9 F cm$^{-1}$ V$^{-1}$ s$^{-1}$ in the saturation region (for $V_{DS}$ between -0.4 and -0.7 V). Furthermore, P(g$_4$2T-T)-based OECTs show a maximum intrinsic gain ($A_i = g_m \times r_o$, with $r_0$ the output resistance $\partial V_{DS}/\partial I_{DS}$) of 62, as reported in Supplementary Figure 5. To further investigate the transient behaviour of the P(g$_4$2T-T)-based OECTs, a transient $I_{DS}$ was measured when a $V_{GS}$ pulse of -0.7V was applied at a fixed $V_{DS}$ = -0.7 V, as shown in Fig. 2c. The P(g$_4$2T-T)-based OECTs exhibit relatively fast switching characteristics with a $\tau_{on}$ of 48 ms and a $\tau_{off}$ of 16 ms while the rise time ($t_r$) is 280 ms and the fall time ($t_f$) is 58 ms (Supplementary Figure 6a). For comparison, P(g$_4$2T-T)-based OECTs fabricated using the same printed electrodes but with an aqueous electrolyte (0.1M NaCl) show an almost identical behaviour to P(g$_4$2T-T)-based OECTs with printed PQ-10 electrolyte (Supplementary Figures 4b, 6b, 7a-c). This indicates that the PQ-10 hydrogel enables the development of printable P(g$_4$2T-T) OECTs without compromising the device performance.

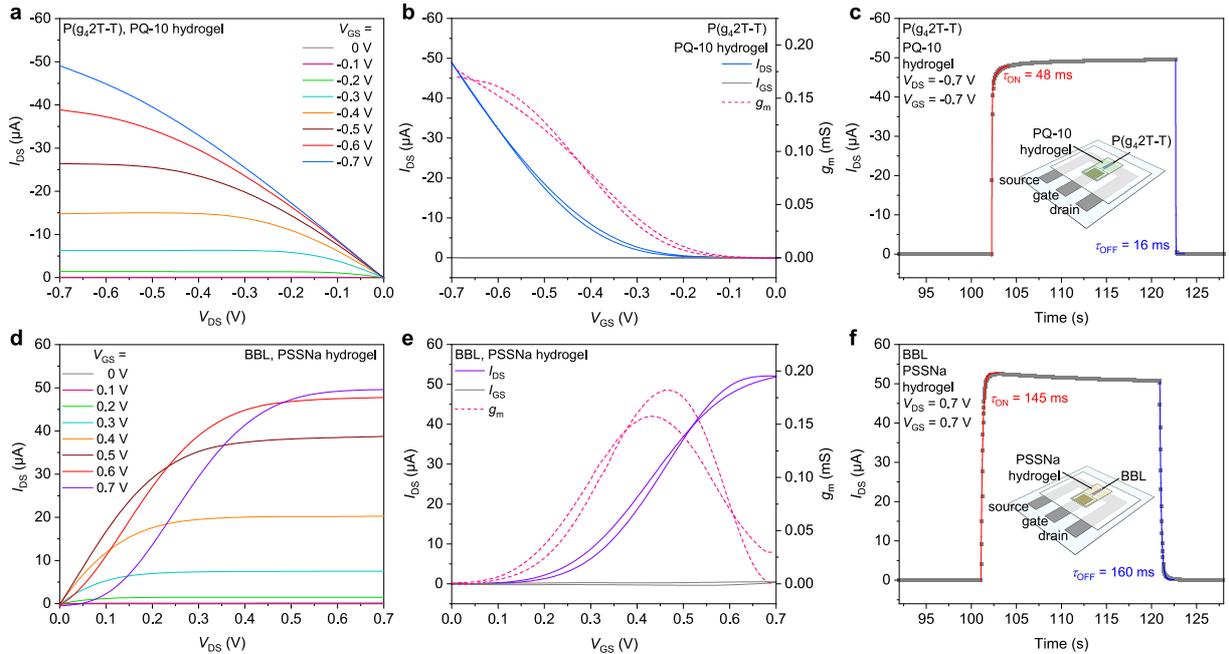

**Fig. 2 | Electrical characteristics of the printed OECTs. a,** Typical output characteristic of printed P(g$_4$2T-T) OECTs. **b,** Transfer characteristics and transconductance in saturation region at $V_{DS}$ = -0.7

V. **c,** The switching characteristics of printed P($g_4$2T-T) OECTs, showing a $\tau_{on}$ of 48 ms and a $\tau_{off}$ of 16 ms. The dimensions of the OECT channel are $W$ = 2000 μm, $L$ = 200 μm, and $d$ = 20 nm. **d,** Typical output characteristic of printed BBL OECTs. **e,** Transfer characteristics and transconductance in saturation region at $V_{DS}$ = -0.7 V. **f,** The switching characteristics of printed BBL OECTs, showing a $\tau_{on}$ of 145 ms and a $\tau_{off}$ of 160 ms. The dimensions of the OECT channel are $W$ = 2000 μm, $L$ = 200 μm, and $d$ = 250 nm.

Robust operation of CMOS-like inverters requires a balance between the driving strengths of the p-type and n-type transistors to maximize noise margins. For instance, the channel width of the PMOS (p-type MOS) transistor in a silicon CMOS inverter is typically advisable to be three times larger than that of the NMOS (n-type MOS) transistor to achieve balanced driving strengths. Because of the volumetric bulk conductivity in OECTs, we are able to achieve a well-balanced p-/n-type OECT operation by using the same channel W and L, but different p-/n-type semiconductor film thickness, so that the same transistor footprint design can be applied regardless of the difference in electron/hole mobility. This OECT feature has no equivalent in other organic and/or inorganic FET technologies, and simplifies substantially the device manufacturing protocol. Fig. 2d-f present the typical electrical characteristics of a printed BBL-based OECT, the characteristics of which were adjusted to match those of the printed P($g_4$2T-T) OECTs in driving strength and operational voltage. As n-type enhancement-mode transistors, a positive voltage on the gate ($V_{GS}$) turns on channel conduction by transporting electrons from source to drain under a positive $V_{DS}$. The ON/OFF current ratio of the printed BBL-based OECTs is typically ~2×10$^3$ and the maximal transconductance is 0.18 mS, as shown in Fig. 2e. It should be noted that the transconductance reaches its maximum at $V_{GS}$ between 0.4 and 0.5 V, rather than at the highest applied $V_{GS}$ = 0.7 V. This suggests that the printed BBL OECTs are optimally operated at $V_{GS}$ between 0.4 and 0.5 V for applications where high transconductance is desirable (e. g., amplifiers). The threshold voltage of the printed BBL-based OECTs is 0.16 V (Supplementary Figure 4), as extracted from the transfer curve in the saturation region (Fig. 2e). Similarly, the $\mu C_V$ of the printed BBL OECTs is calculated to be 2.63 F cm$^{-1}$ V$^{-1}$ s$^{-1}$, using Eq. (2). A maximum intrinsic gain $A_i$ of 63 is shown in Supplementary Figure 5. Fig. 2f shows the switching characteristics of the printed BBL OECTs with a $\tau_{on}$ of 145 ms and a $\tau_{off}$ of 160 ms. The corresponding $t_r$ and $t_f$ are 390 ms and 360 ms, respectively (Supplementary Figure 6c). For comparison, BBL-based OECTs with a 0.1 M NaCl aqueous electrolyte show similar characteristics (Supplementary Figure 4d, 6d, 7d-f). However, the maximum transconductance of BBL-based OECTs with NaCl electrolyte is shifted to slightly larger $V_{GS}$ = 0.6 V, suggesting that printed BBL OECTs with PSSNa hydrogel are more favourably operated at lower voltages.

**In-situ spectroelectrochemistry of OECTs**
To understand the operation of P($g_4$2T-T) and BBL OECTs, we measured absorption spectra of the polymer films on FTO inside a three-electrode electrochemical cell. We used 0.1 M NaCl as the electrolyte and an Ag/AgCl reference electrode to match the OECT characterization. The absorption spectra of the pristine polymers in air are presented in Supplementary Figure 8. Pristine P(g$_4$2T-T) films have an absorption maximum at 600 nm followed by a broad polaronic absorption extending into the IR range due to oxygen doping of P(g$_4$2T-T) in air.[25] Pristine BBL on the other hand has an absorption maximum at 570 nm with no visible IR tail due to its undoped state. For spectroelectrochemical measurements, the polymers were first fully dedoped by applying $V_{GS}$ = 0.5 V for P(g$_4$2T-T) and $V_{GS}$ = -0.7 V for BBL, followed by measuring the baseline. The application of negative gate voltages for P(g$_4$2T-T) and positive gate voltages for BBL results in the injection of charges into the polymer layer, compensated by ion diffusion from the electrolyte. The differential absorption spectra from the undoped state during the CV scans are presented in Fig 3a-c for P(g$_4$2T-T) and in Fig 3d-f for BBL. The formation of polarons is clearly seen in the absorption spectra as a bleaching of the ground state absorption band and an accompanying formation of polaronic absorption for both polymers.

For P(g$_4$2T-T) the bleaching of the ground state is especially notable, as the extent of the bleaching is larger in magnitude than the ground state absorption of the same sample measured in air. It is important to note that since P(g$_4$2T-T) is doped in air, the undoped baseline in the spectroelectrochemistry measurements has a larger ground state absorption. Nevertheless, most of the P(g$_4$2T-T) ground state absorption is bleached, illustrating that a majority of P(g$_4$2T-T) segments are doped at the highest applied voltages. The bleaching is accompanied by a broad polaronic absorption band extending well into the IR region. The polaronic absorption reaches a maximum at $V_{GS}$ = -0.3 V, followed by a decrease in the polaronic absorption while the ground state further bleaches slightly at more negative potentials. This is attributed to increased bipolaron formation at more oxidative potentials, as recently shown for P(g$_4$2T-T).[26,27]

Unlike P(g$_4$2T-T), the bleaching of BBL ground state absorption does not have the same shape as the pristine ground state absorption shown in Supplementary Figure 8. This is due to overlapping positive polaronic absorption bands, indicating that the fraction of polaronic BBL cannot be directly derived from the amplitude of the ground state bleach. In addition, BBL has multiple positive polaronic absorption bands with maxima at 400, 720, and 865 nm. The 720 and 865 nm bands are close to each other in energy and form one broad absorption band at the largest applied gate voltages. This band reaches a magnitude that is a third in intensity of the pristine ground state absorption, which is more intense than previously published spectra of molecularly-doped BBL films.[25] We can conclusively say that the bulk of both Pg($_4$2T-T) and BBL are extensively doped under operating conditions, contrasting to what would be expected if the materials were operating in a field effect regime, where only a thin layer at the polymer surface forms a conductive path.

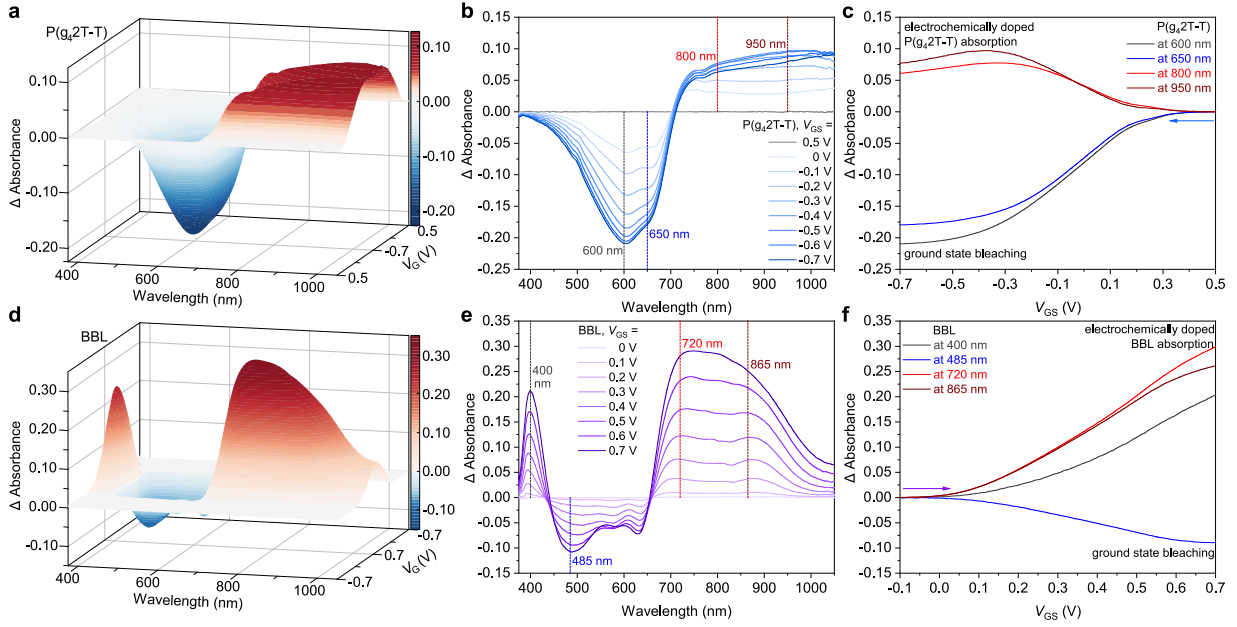

**Fig. 3 | Differential spectroelectrochemistry**. **a,** Differential absorption spectra of P(g$_4$2T-T) scanned between 0.5 V and -0.7 V vs. Ag/AgCl. **b,** Differential absorption spectra of P(g$_4$2T-T) at select voltages during the negative CV scan. **c,** Differential absorbance of P(g$_4$2T-T) at select wavelengths during the negative CV scan. **d,** Differential absorption spectra of BBL scanned between -0.7 V and 0.7 V vs. Ag/AgCl. **e,** Differential absorption spectra of BBL at select voltages during the positive CV scan. **f,** Differential absorbance of BBL at select wavelengths during the positive CV scan. All of the absorption values are reported as difference absorbance from the undoped state, the voltage scan rate is 50 mV/s.

**Printed OECT-based complementary inverters**

Having validated a pair of well-balanced p-type and n-type OECTs working in enhancement-mode, we demonstrate a printed complementary inverter that yields large gains (~26 V/V) at a low supply voltage (< 0.7 V). The schematic of an OECT-based complementary inverter is shown in Fig. 4a where P(g$_4$2T-T) and BBL OECTs are switched either ON or OFF depending on the input voltage $V_{in}$ at the gate, resulting in a $V_{out}$ which is either equal to $V_{DD}$ (ON, p-type doped, n-type dedoped) or to the ground (OFF, n-type doped, p-type dedoped). To make the transistors switch properly, their $V_T$ play an important role. The low $V_T$ for both P(g$_4$2T-T) and BBL OECTs enables the low voltage operation of the printed complementary inverter.

The switching threshold of an inverter ($V_M$), defined as when $V_{in}$ equals $V_{out}$, is an indication of the balanced driving strengths of p-type and n-type transistors, and ideally, the threshold is equal to $V_{DD}/2$. The switching threshold is shown in Fig. 4b, and it closely follows the ideal behaviour at all $V_{DD}$. Another benefit of balanced p-type and n-type OECTs is to obtain large *high* and *low* noise margins ($NM_H$ and $NM_L$), defined as follows:

$$NM_H = V_{DD} - V_{IH}, NM_L = V_{IL}, \tag{3}$$

where $V_{IH}$ and $V_{IL}$ are the input voltages HIGH and LOW at the operation points of $\frac{\partial V_{out}}{\partial V_{in}} = -1$ at the voltage transfer characteristics (VTC*)* of the inverter (Fig. 5b). Here, the $NM_H$ and $NM_L$ are listed in Supplementary Table 1 and the total *NM* with respect to $V_{DD}$ is up to 89 % ($V_{DD}$ = 0.7 V).

The voltage gain $A_v$ in units of V/V and decibels (dB) is defined as:

$$A_V = \left|\frac{\partial V_{out}}{\partial V_{in}}\right| \text{ (V/V)}, \quad A_V = 20 log \left|\frac{\partial V_{out}}{\partial V_{in}}\right| \text{ (dB)}. \tag{4}$$

As shown in Fig. 4c, the maximum gain reaches 26 V/V at $V_{DD}$ = 0.7 V, and a relatively high gain of 7.5 V/V can be obtained even with a very low supply voltage of 0.3 V. When $V_{DD}$ is further reduced to 0.2 V, a major deterioration in the VTC is observed with a greatly reduced gain of 2.7 V/V. This is due to the supply voltage approaching the threshold voltage limit. For comparison, unipolar inverters composed of the same printed OECTs (either n-type or p-type only) show maximum gain lower than 1.8 V/V even for $V_{DD}$ of 0.7 V (Supplementary Figure 9 and 10).

Low power consumption is another major advantage of CMOS-like technologies since one of the two transistors is always in the non-conducting OFF-state in static conditions. The current at $V_{DD}$ = 0.7 V is below 17 nA, yielding a static power consumption of only 12 nW, which is far lower than that of state-of-the-art unipolar OECT-based inverters (1.18 mW)[6]. Even during the switching of the inverter, the maximum power consumption per logic gate is just 2.7 µW, and reaches only 0.10 µW at lower supply voltage (0.3 V). For comparison, previous attempts to develop a complementary OECT-based inverter[20] yielded a maximum power consumption of 15.1 µW and a static power consumption of 1.35 µW at $V_{DD}$ = 0.6 V. If implemented in fully-printed large-scale digital circuits, such as a 4-to-7 decoder and 7-bit shift register[6], the complementary OECTs in this work could enable a power consumption reduction of up to 99.8%.

In order to thoroughly evaluate the performance of the inverters, a five-stage ring oscillator, the largest circuit based on complementary OECTs in terms of the number of OECTs, was implemented by cascading five inverters (Supplementary Figures 11 and 12). This OECT-based ring oscillator shows a stable oscillation with a period of 6.9 s which gives a stage delay of 690 ms by $t = 1/(2Nf)$, where *f* is the oscillation frequency and *N* is the number of stages. The stage delay is consistent with the switching characteristics of the individual OECTs and shows the true operation speed in multi-stage circuitries.

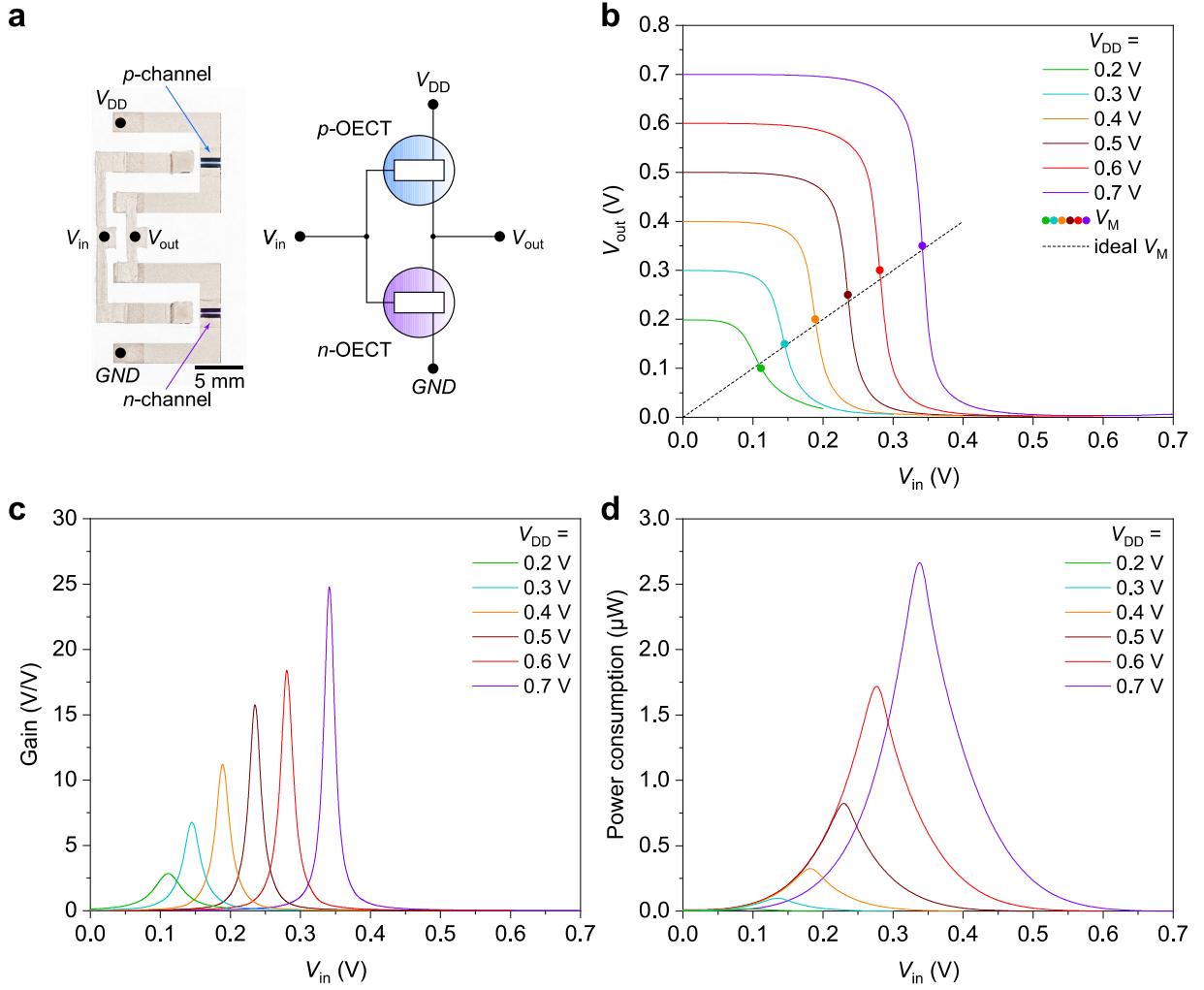

**Fig. 4 | Printed OECT-based complementary inverters. a,** Photograph and schematic of a printed complementary inverter. **b,** Typical voltage transfer characteristics (*VTC*) of the printed inverter at various supply voltages (0.7 to 0.2 V). **c,** Voltage gain of the inverter at various supply voltages. **d,** Power consumption of the inverter at various supply voltages.

**Printed OECT-based push-pull voltage amplifiers**

Although typically implemented as digital devices, CMOS inverters can be used as analog amplifiers. A major advantage of using CMOS inverters as amplifiers in the push-pull configuration (Fig. 5a), as opposed to active load and/or current source load configurations, is that these devices benefit from the summation of the transconductance of both transistors[21], as shown in Eq. (5). The voltage gain of a OECT-based CMOS-like inverter can be calculated as:

$$A_V = -(g_{mp} + g_{mn})(r_{op}/r_{on}), \tag{5}$$

where $g_{mp}$ and $g_{mn}$ are the transconductances and $r_{op}$ and $r_{on}$ are output resistances of the p-type and n-type transistors, respectively. The main challenge in operating OECT-based CMOS-like inverters is to correctly bias the inverter to position the operation point in the transition region (close to the switching threshold $V_M$), where both transistors work in the saturation region and the inverter exhibits a large gain.

Our push-pull amplifiers with printed complementary OECTs show a gain of 16 (V/V) (24 dB) when amplifying a 10 mV (amplitude) sinusoidal signal with a DC offset of 0.28 V (Fig. 5b), which is consistent with the DC gain of the inverter shown in Fig. 4c. When amplifying signals as low as 100 µV, the voltage gain reaches 29 (V/V) (29.2 dB), as shown in Supplementary Figures 13-14. The maximum amplification observed is 33 (V/V) (30.4 dB) with a 200 µV input signal (Fig. 5c). We evaluated the gain of the voltage amplifier with respect to input frequencies ranging from 0.06 Hz to 5 Hz with a 10 mV input signal. The detailed waveforms at each frequency are shown in Supplementary Figures 13 and 14.

These OECT-based single stage push-pull amplifier can serve as a building block for more complex amplifiers. Here, we demonstrate a two-stage amplifier to obtain a high voltage gain by cascading a number of single-stage amplifiers, as shown in Fig. 5d. The two-stage amplifier exhibits a voltage gain as high as 193 (V/V), which is the highest reported for OECT-based amplifiers. For comparison, the highest voltage gain among OECT technologies reported to date is 107 (V/V) for depletion-mode, p-type PEDOT:PSS-based OECTs. These unipolar OECTs however require a supply voltage of 0.8 V and consume ~480 µW of power[19]. Here, we are able to almost double the voltage amplification while reducing the maximum power consumption (5.6 µW) by close to 99%.

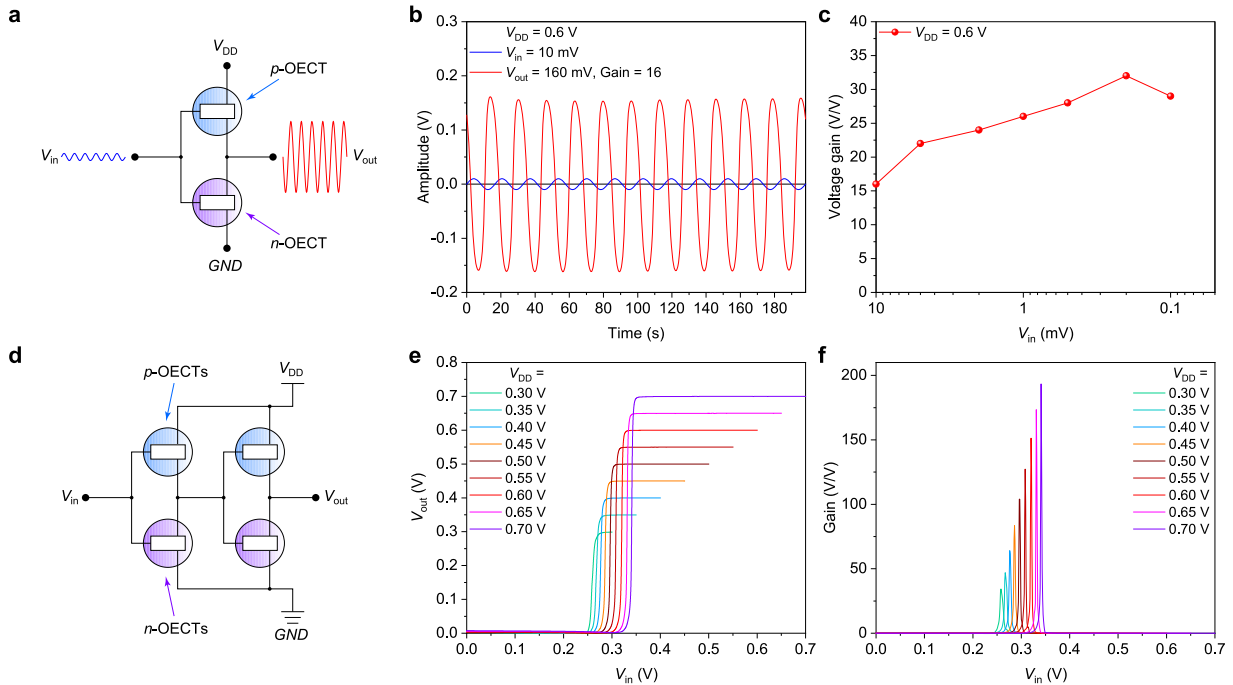

**Fig. 5 | Printed OECT-based pull-push voltage amplifier. a,** Illustration of the amplifier configuration used to obtain these measurements. **b,** Voltage waveform of input and output signals when a 10 mV and 0.06 Hz sinusoid voltage is applied as the input. **c,** Voltage gain dependence on the amplitude of the input signals (10mV to 100µV). **d,** Schematic of a five-stage pull-push CMOS amplifier. **e,** The voltage transfer characteristics (VTC) of the five-stage amplifier. **f,** Voltage gain of the five-stage amplifier.

By coupling the printed enhancement-mode OECTs operated at low voltage and the complementary push-pull configuration, we are able to demonstrate flexible voltage amplifiers with high gain and very low power dissipation. In order to evaluate the amplifier performance with respect to power efficiency, we normalize the voltage gain to the maximum power consumption at various supply voltages using the equation $A_{V/P} = A_V/P_{max}$ (dB/µW). Here, the normalized voltage gain $A_{V/P}$ for the printed voltage amplifier (single stage) reaches 10.2 dB/µW at $V_{DD}$ = 0.7 V and 169 dB/µW at $V_{DD}$ = 0.3 V, which is up to 50 times higher than that of state-of-the-art OECT amplifiers[17–19] (Fig. 6a). In addition to existing OECT technologies, we have also compared our results with voltage amplifiers based on electrolyte-gated thin-film transistors[28–31] and organic field-effect transistors[32–36] (details can be found in Supplementary Table 2). Cutting-edge voltage amplifiers, that implement unipolar organic field-effect transistors in differential[36] and pseudo-CMOS[35] circuit configurations and are typically operated at a few volts, reach normalized gain values of 5 dB/µW (differential) and 4.6 dB/µW (pseudo-CMOS). Electrolyte-gated thin-film transistors (e.g., IGZO[30], CNT[31]) are usually operated at around 1 V and generally exhibit normalized gains between 0.03 and 0.75 dB/µW. From the comparison, we can see that complementary OECT-based amplifiers offer a power-efficient solution to sense weak voltage signals (as low as 100 µV) at low supply voltages (as low as 0.3 V). Power efficiency is crucial for self-powered applications[37] where the power supply is heavily limited. For example, commercially available printed, flexible organic solar cells can yield about 18 µW/cm$^2$ with indoor lighting[38]. Thermoelectric generators typically offer a few microvolts with a maximun of 28.5 µW/cm$^2$ for a temperature difference of 10 °C[39]. Electrostatic generators can provide around 50 µW/cm$^2$ [40], while electromagnetic generators can supply up to 150 µW/cm$^2$ [41]. Assuming that a self-powered device has an area of 1 cm$^2$ dedicated to carry the power supplier, the power that can be provided by such energy supplier is shown in Fig. 6a. This shows that our printed voltage amplifier has huge potential for self-powered devices.

As we demonstrated, the voltage gain of the printed amplifier can be further increased by cascading multiple stages together. With two-stage printed amplifiers, we report DC gains of 193 V/V, which, to the best of our knowledge, are the highest not only among the OECT technologies[19,20] but also among other sub-1V amplifiers based on emerging CMOS-like technologies like CNTs[42,43], graphene[44], and 2D transition metal dichalcogenide[45–50] (see Fig. 6b and Supplementary Table 3). As an example, a hybrid CMOS-like inverter based on CNT and quantum dot transistors operates with a DC gain of 76 V/V at $V_{DD}$ = 0.9 V, while the highest gain among 2D transition metal dichalcogenide transistors (p-type WSe$_2$ and n-type MoS$_2$) is 34 V/V at $V_{DD}$ = 1 V.

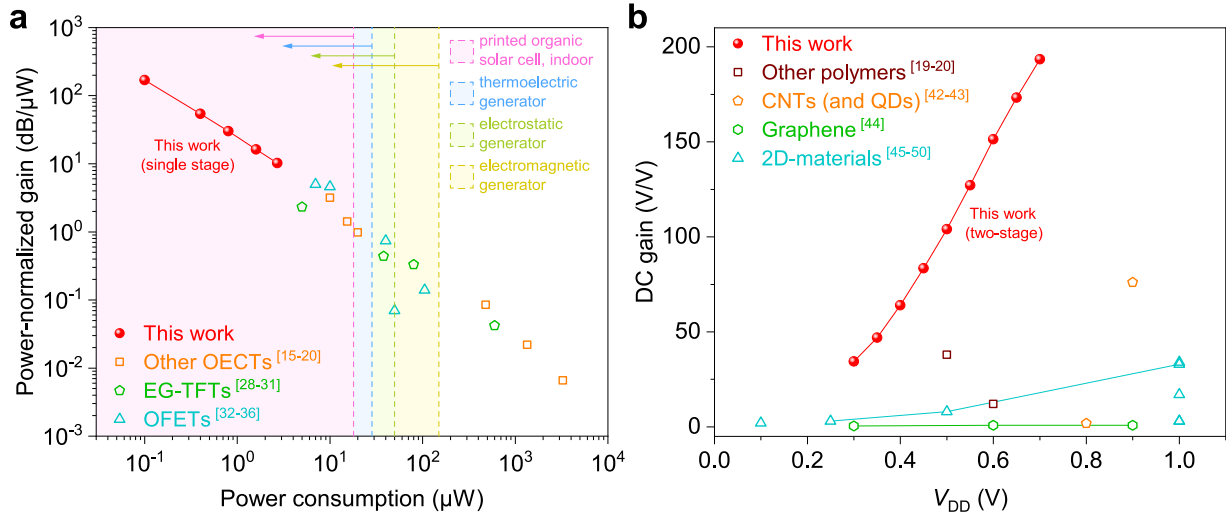

**Fig. 6 | Power efficiency and voltage gain benchmarks. a,** Power efficiency benchmarks of voltage amplifiers based on OECTs, electrolyte-gated thin-film transistors, and organic field effect transistors. **b,** DC voltage gain (V/V) among reported sub-1V CMOS-like technologies.

**Conclusions**

We have developed flexible, printed digital and analog circuits composed of p-/n-type enhancement mode OECTs. The p- and n-type OECTs incorporate P($g_4$2T-T) and BBL as the channel material, and polyanionic PQ-10 and polycationic PSSNa hydrogel as the electrolyte, respectively. By engineering the thickness of the active materials, we match the driving strengths of the p-type and n-type OECTs with identical channel dimensions. Both types of OECTs exhibit low $V_T$ (-0.22 V for P($g_4$2T-T) OECTs and 0.16 V for BBL OECTs), which enables the single-stage amplifier to operate at a low voltage and with a power-normalized gain of up to 169 dB/µW, which is 53 times higher than current state-of-the-art OECT amplifiers. Furthermore, the two-stage amplifier reaches a DC gain of 193 V/V, which is a record-high value among all emerging sub-1V CMOS-like technologies.

With such a high power efficiency, we envisage that our flexible and fully-printable OECT-based voltage amplifiers may find several applications in recording/monitoring of weak voltage signals, such as IoT sensing, electrophysiological and other bioelectric infra-slow signals[51,52], with limited availability to power supply or a need for on-site amplification. The progress of electrophysiological signal recordings and electronic neural interfaces has been tending towards autonomous, implantable, wearable devices integrated with sensing electrodes and signal processing units[53]. Long-term wireless monitoring on a mobile curvilinear surface requires flexible and power-efficient amplifiers that operate at low voltages and draw little current to extend the limits imposed by battery capacity. Similar requirements apply also to IoT applications, where scavenging and storage of energy suggest low-power amplification and A/D conversion and processing on the module. To summarize, our printed complementary voltage amplifiers may address many key the challenges posed in the wireless detection of weak sensor signals by demonstrating the potentials of a) in-situ pre-amplification of small signals, b) flexibility for conformables/wearables, c) low power consumption for implants, d) scalability for massively parallel recording by screen-printing manufacturing, e) electrolyte nature at the input (gates of OECTs) for good interface to skin and organs, and f) potential signal filtering and processing with OECT-based circuitries.

**Methods**

**Materials.** The polyethylene terephtalate (PET) substrate *Polifoil Bias* is purchased from *Policrom Screen*. Silver ink (*Ag 5000, DuPont*) is used for printed interconnects. Carbon ink *7102* printing paste (*DuPont*) is used for drain/source/gate contacts. Insulating ink (*5018, DuPont*) is used for electrode isolation. PQ-10, PSSNa, and BBL were purchased from Sigma-Aldrich and used as received. P(g$_4$2T-T) was synthesized following the procedure reported previously ($M_n$ = 24 kDa, polydispersity index = 3.3).[54]

**Sample preparation.** The P(g$_4$2T-T) and BBL inks were fabricated through a surfactant-free method. P(g$_4$2T-T)(30 mg) was dissolved in chloroform (15 mL) to form a dark magenta solution, then the P(g$_4$2T-T)- chloroform solution was dropwise added to isopropanol (75 mL) under high speed stirring (1500 rpm). During the solvent-exchange, bright blue nanoparticles were generated. BBL (30 mg) was dissolved in MSA (15 mL) to form a deep red solution, then the BBL-MSA solution was dropwise added to isopropanol (75 mL) under high speed stirring (1500 rpm). During the solvent-exchange, dark purple (for BBL) nanoparticles were generated. The P(g$_4$2T-T) and BBL nanoparticles were respectively collected by centrifugation (5000 rpm, 30 min) and washed by IPA for 6 times until neutral. The neutral P(g$_4$2T-T) and BBL nanoparticles were re-dispersed in IPA to obtain a dispersion inks (about 0.006 mg/mL for P(g$_4$2T-T) and 0.1 mg/mL for BBL). The preparation of the electrolyte hydrogels was done as follows. PQ-10 (2.5 g) was added to water (7.5 mL) in a 20 mL reaction vessel and was then sealed and stirred at 150 °C for 4 hours until a uniform, bubble-free highly viscous hydrogel was formed. The PQ-10 hydrogel was then cooled down to room temperature before use. PSSNa (4.0 g), *d*-sorbitol (1.0 g), and glycerol (1.0 g) were added to water (4.0 mL) in a 20 mL reaction vessel and the mixer was sealed and stirred at 150 °C for 4 hours until a uniform, bubble-free highly viscous hydrogel was formed. The PSSNa hydrogel was then cooled down to room temperature before use.

**Fabrication of OECTs and inverters/amplifiers.** All materials were deposited by flatbed sheet-fed screen printing equipment (*DEK Horizon 03iX*) on top of PET plastic substrates. The printed features of the design layout cover approximately the area of an A4 sheet. A layer of carbon is first deposited on the substrate to form the contacts. *Ag 5000* silver ink is then patterned, followed by an insulating layer of *5018* ink, which is UV-cured to define the area of the gate and channel. Ag/AgCl gate layer was fabricated by blade-coating using Ag/AgCl ink through a shadow mask, then annealed at 100 °C for 2 min. The P(g$_4$2T-T) or BBL layer were fabricated by spray-casting in air through a shadow mask, by means of a standard HD-130 air-brush (0.3mm) with atomization air pressure of 1 bar. The PQ-10 or PSSNa hydrogel electrolyte layer was fabricated by blade-coating through a shadow mask.

**Spectroelectrochemistry.** Absorption measurements were performed on BBL and P(g$_4$2T-T) electrodes on FTO glass using an AvaSpec ULS2048L fiber optic spectrometer and an AvaLight HAL-S-Mini light source (Avantes BV). The electrode was immersed in 0.1 M NaCl solution inside a three-electrode electrochemical cell with a platinum wire counter electrode and a Ag/AgCl reference electrode. CV scans were performed with a potentiostat (BioLogic SP-200). The electrodes were first fully dedoped by applying a potential of -0.7 V for BBL and 0.5 V for P(g$_4$2T-T) for 10 s, during which time the absorption baseline was measured, followed by running a CV scan between -0.7 V and 0.7 V for BBL and between 0.5 V and -0.7 V for P(g$_4$2T-T) to match with the gate voltage range applied to the transistors. The scan speed was 50 mV/s. The potentiostat triggered the absorption measurement to begin at the exact same time as the CV scan, with averages of 8 absorption spectra saved every 10 ms.

**Characterization of OECTs and amplifiers.** All measurements are performed in a controlled environment at a temperature of ~20 °C and at a relative humidity of ~45 %. Transfer and dynamic switch measurements were performed by a semiconductor parameter analyser (*Keithley 4200 SCS*) and a probe station.

**Data availability**

The data that support the findings of this study are available from the corresponding author upon request.

**References**

1. Dennard, R. H. *et al.* Design of Ion-Implanted MOSFET's With Very Small Physical Dimensions. *IEEE J. Solid-State Circuits* **9**, 256–268 (1974).
2. Hiramoto, T. Five nanometre CMOS technology. *Nat. Electron.* **2**, 557–558 (2019).
3. De Donno, D., Catarinucci, L. & Tarricone, L. RAMSES: RFID augmented module for smart environmental sensing. *IEEE Trans. Instrum. Meas.* **63**, 1701–1708 (2014).
4. Rivnay, J. *et al.* Organic electrochemical transistors. *Nat. Rev. Mater.* **3**, 1–14 (2018).
5. Tu, D. & Fabiano, S. Mixed ion-electron transport in organic electrochemical transistors. *Appl. Phys. Lett.* **117**, 080501 (2020).
6. Andersson Ersman, P. *et al.* All-printed large-scale integrated circuits based on organic electrochemical transistors. *Nat. Commun.* **10**, 5053 (2019).
7. Andersson Ersman, P. *et al.* Screen printed digital circuits based on vertical organic electrochemical transistors. *Flex. Print. Electron.* **2**, 045008 (2017).
8. Hutter, P. C., Rothlander, T., Scheipl, G. & Stadlober, B. All Screen-Printed Logic Gates Based on Organic Electrochemical Transistors. *IEEE Trans. Electron Devices* **62**, 4231–4236 (2015).
9. Someya, T., Bao, Z. & Malliaras, G. G. The rise of plastic bioelectronics. *Nature* **540**, 379–385 (2016).

10. Berggren, M. & Richter-Dahlfors, A. Organic Bioelectronics. *Adv. Mater.* **19**, 3201–3213 (2007).
11. Simon, D. T., Gabrielsson, E. O., Tybrandt, K. & Berggren, M. Organic Bioelectronics: Bridging the Signaling Gap between Biology and Technology. *Chem. Rev.* **116**, 13009–13041 (2016).
12. Gerasimov, J. Y. *et al.* An Evolvable Organic Electrochemical Transistor for Neuromorphic Applications. *Adv. Sci.* **6**, 1801339 (2019).
13. Gkoupidenis, P., Koutsouras, D. A. & Malliaras, G. G. Neuromorphic device architectures with global connectivity through electrolyte gating. *Nat. Commun.* **8**, 1–8 (2017).
14. Svensson, P. O., Nilsson, D., Forchheimer, R. & Berggren, M. A sensor circuit using reference-based conductance switching in organic electrochemical transistors. *Appl. Phys. Lett.* **93**, 203301 (2008).
15. Rivnay, J. *et al.* Organic electrochemical transistors with maximum transconductance at zero gate bias. *Adv. Mater.* **25**, 7010–7014 (2013).
16. Braendlein, M., Lonjaret, T., Leleux, P., Badier, J.-M. & Malliaras, G. G. Voltage Amplifier Based on Organic Electrochemical Transistor. *Adv. Sci.* **4**, 1600247 (2017).
17. Venkatraman, V. *et al.* Subthreshold Operation of Organic Electrochemical Transistors for Biosignal Amplification. *Adv. Sci.* **5**, 1800453 (2018).
18. Romele, P., Ghittorelli, M., Kovács-Vajna, Z. M. & Torricelli, F. Ion buffering and interface charge enable high performance electronics with organic electrochemical transistors. *Nat. Commun.* **10**, 1–11 (2019).
19. Romele, P. *et al.* Multiscale real time and high sensitivity ion detection with complementary organic electrochemical transistors amplifier. *Nat. Commun.* **11**, 3743 (2020).
20. Sun, H. *et al.* Complementary Logic Circuits Based on High-Performance n-Type Organic Electrochemical Transistors. *Adv. Mater.* **30**, 1704916 (2018).
21. Sharroush, S. M. Design of the CMOS inverter-based amplifier: A quantitative approach. *Int. J. Circuit Theory Appl.* **47**, 1006–1036 (2019).
22. Ersman, P. A. *et al.* Monolithic integration of display driver circuits and displays manufactured by screen printing. *Flex. Print. Electron.* **5**, 024001 (2020).
23. Schmatz, B., Lang, A. W. & Reynolds, J. R. Fully Printed Organic Electrochemical Transistors from Green Solvents. *Adv. Funct. Mater.* **29**, 1905266 (2019).
24. Keene, S. T. *et al.* Enhancement-Mode PEDOT:PSS Organic Electrochemical Transistors Using Molecular De-Doping. *Adv. Mater.* **32**, 2000270 (2020).
25. Xu, K. *et al.* Ground-state electron transfer in all-polymer donor–acceptor heterojunctions. *Nat. Mater.* **19**, 738–744 (2020).

26. Sahalianov, I. *et al.* UV-to-IR Absorption of Molecularly p-Doped Polythiophenes with Alkyl and Oligoether Side Chains: Experiment and Interpretation Based on Density Functional Theory. *J. Phys. Chem. B* **124**, 11280–11293 (2020).

27. Zokaei, S. *et al.* Toughening of a Soft Polar Polythiophene through Copolymerization with Hard Urethane Segments. *Adv. Sci.* **8**, 2002778 (2021).

28. Xia, Y. *et al.* Printed sub-2 V Gel-electrolyte-gated polymer transistors and circuits. *Adv. Funct. Mater.* **20**, 587–594 (2010).

29. Cho, K. G. *et al.* Sub-2 V, Transfer-Stamped Organic/Inorganic Complementary Inverters Based on Electrolyte-Gated Transistors. *ACS Appl. Mater. Interfaces* **10**, 40672–40680 (2018).

30. Park, S. *et al.* Sub-0.5 v Highly Stable Aqueous Salt Gated Metal Oxide Electronics. *Sci. Rep.* **5**, 1–9 (2015).

31. Ha, M. *et al.* Aerosol Jet Printed, Low Voltage, Electrolyte Gated Carbon Nanotube Ring Oscillators with Sub-5 μs Stage Delays. *Nano Lett.* **13**, 954–960 (2013).

32. Guerin, M. *et al.* High-gain fully printed organic complementary circuits on flexible plastic foils. *IEEE Trans. Electron Devices* **58**, 3587–3593 (2011).

33. Marien, H., Steyaert, M. S. J., Van Veenendaal, E. & Heremans, P. A fully integrated ΔΣ ADC in organic thin-film transistor technology on flexible plastic foil. *IEEE J. Solid-State Circuits* **46**, 276–284 (2011).

34. Fukuda, K. *et al.* Fully-printed high-performance organic thin-film transistors and circuitry on one-micron-thick polymer films. *Nat. Commun.* **5**, 1–8 (2014).

35. Sekitani, T. *et al.* Ultraflexible organic amplifier with biocompatible gel electrodes. *Nat. Commun.* **7**, 1–11 (2016).

36. Sugiyama, M. *et al.* An ultraflexible organic differential amplifier for recording electrocardiograms. *Nat. Electron.* **2**, 351–360 (2019).

37. Park, S. *et al.* Self-powered ultra-flexible electronics via nano-grating-patterned organic photovoltaics. *Nature* **561**, 516–521 (2018).

38. https://www.epishine.com.

39. Amar, A. Ben, Kouki, A. B. & Cao, H. Power approaches for implantable medical devices. *Sensors (Switzerland)* **15**, 28889–28914 (2015).

40. Tashiro, R., Kabei, N., Katayama, K., Tsuboi, F. & Tsuchiya, K. Development of an electrostatic generator for a cardiac pacemaker that harnesses the ventricular wall motion. *J. Artif. Organs* **5**, 239–245 (2002).

41. Amirtharajah, R. & Chandrakasan, A. P. Self-powered signal processing using vibration-based power generation. *IEEE J. Solid-State Circuits* **33**, 687–695 (1998).

42. Joshi, S. *et al.* Ambient Processed, Water-Stable, Aqueous-Gated sub 1 V n-type Carbon Nanotube Field Effect Transistor. *Sci. Rep.* **8**, 11386 (2018).

43. Shulga, A. G. *et al.* An All-Solution-Based Hybrid CMOS-Like Quantum Dot/Carbon Nanotube Inverter. *Adv. Mater.* **29**, 1701764 (2017).

44. Dathbun, A., Kim, S., Lee, S., Hwang, D. K. & Cho, J. H. Flexible and transparent graphene complementary logic gates. *Mol. Syst. Des. Eng.* **4**, 484–490 (2019).

45. Alam, M. H. *et al.* Lithium-ion electrolytic substrates for sub-1V high-performance transition metal dichalcogenide transistors and amplifiers. *Nat. Commun.* (2020) doi:10.1038/s41467-020-17006-w.

46. Pezeshki, A. *et al.* Static and dynamic performance of complementary inverters based on nanosheet α-MoTe2 p-channel and MoS2 n-channel transistors. *ACS Nano* **10**, 1118–1125 (2016).

47. Lim, J. Y. *et al.* Homogeneous 2D MoTe$_2$ p-n Junctions and CMOS Inverters formed by Atomic-Layer-Deposition-Induced Doping. *Adv. Mater.* **29**, 1701798 (2017).

48. Lan, Y. W. *et al.* Scalable fabrication of a complementary logic inverter based on MoS2 fin-shaped field effect transistors. *Nanoscale Horizons* **4**, 683–688 (2019).

49. Tosun, M. *et al.* High-gain inverters based on WSe2 complementary field-effect transistors. *ACS Nano* **8**, 4948–4953 (2014).

50. Larentis, S. *et al.* Reconfigurable Complementary Monolayer MoTe2 Field-Effect Transistors for Integrated Circuits. *ACS Nano* **11**, 4832–4839 (2017).

51. Lázár, Z. I., Dijk, D. J. & Lázár, A. S. Infraslow oscillations in human sleep spindle activity. *J. Neurosci. Methods* **316**, 22–34 (2019).

52. Masvidal-Codina, E. *et al.* High-resolution mapping of infraslow cortical brain activity enabled by graphene microtransistors. *Nat. Mater.* **18**, 280–288 (2019).

53. Zhang, M., Tang, Z., Liu, X. & Van der Spiegel, J. Electronic neural interfaces. *Nat. Electron.* **3**, 191–200 (2020).

54. Kroon, R. *et al.* Polar Side Chains Enhance Processability, Electrical Conductivity, and Thermal Stability of a Molecularly p-Doped Polythiophene. *Adv. Mater.* **29**, 1700930 (2017).

**Acknowledgements**

The authors thank Prof. Robert Forchheimer (Linköping U.), Prof. Kai Xu (Yanshan U.), Prof. Hengda Sun (Donghua U.), Dr. Mary Donahue (Linköping U.) and Silan Zhang (Linköping U.) for helpful discussion. We also thank Marie Nilsson and Jan Strandberg (RISE) for assistance with screen-printing the electrodes. This work was financially supported by the Knut and Alice Wallenberg foundation, the Swedish Research Council (2016-03979 and 2020-03243), Swedish Foundation for Strategic Research (SE13-0045), ÅForsk (18-313 and 19-310), Olle Engkvists Stiftelse (204-0256), VINNOVA (2020-05223), and the Swedish Government Strategic Research Area in Materials Science on Functional Materials at Linköping University (Faculty Grant SFO-Mat-LiU 2009-00971).

**Author contributions**

S.F. conceived and designed the experiments. C.-Y.Y. developed the P(g$_4$2T-T)/BBL inks and the PSSNa/PQ-10 hydrogels. D.T. and J.Y.G designed the layout for printing, C.-Y.Y. fabricated the OECTs, inverters and ring oscilators, C.-Y.Y. and D.T. performed the electrical characterization measurements and analysed the data. P.C.H. helped with the OECT fabrication. T.-P.R. recorded and analysed the UV–Vis–NIR data. R.K. and C.M. synthesized P(g$_4$2T-T). H.-Y.W. measured the ions conductivity. D.T., C.-Y.Y., T.-P.R., J.Y.G., M.B., and S.F. wrote the manuscript. All authors contributed to discussion and manuscript preparation.

**Competing interests**

The authors declare that they have no competing interests.

**Additional information**

The authors declare no competing interests.

**Correspondence and requests for materials** should be addressed to S.F.

*Supplementary Information*

# Low-power/high-gain flexible complementary circuits based on printed organic electrochemical transistors

Chi-Yuan Yang[1], Deyu Tu[1], Tero-Petri Ruoko[1], Jennifer Y. Gerasimov[1], Han-Yan Wu[1], P. C. Harikesh[1], Renee Kroon[1,2], Christian Müller[3], Magnus Berggren[1,2,4], Simone Fabiano[1,2,4]*

[1]Laboratory of Organic Electronics, Department of Science and Technology, Linköping University, SE-601 74 Norrköping, Sweden.

[2]Wallenberg Wood Science Center, Linköping University, SE-601 74 Norrköping, Sweden.

[3]Wallenberg Wood Science Center, Department of Chemistry and Chemical Engineering, Chalmers University of Technology, SE-412 06 Göteborg, Sweden.

[4]n-Ink AB, Teknikringen 7, SE-583 30 Linköping, Sweden.

Correspondence should be addressed to: simone.fabiano@liu.se

**Supplementary Figures**

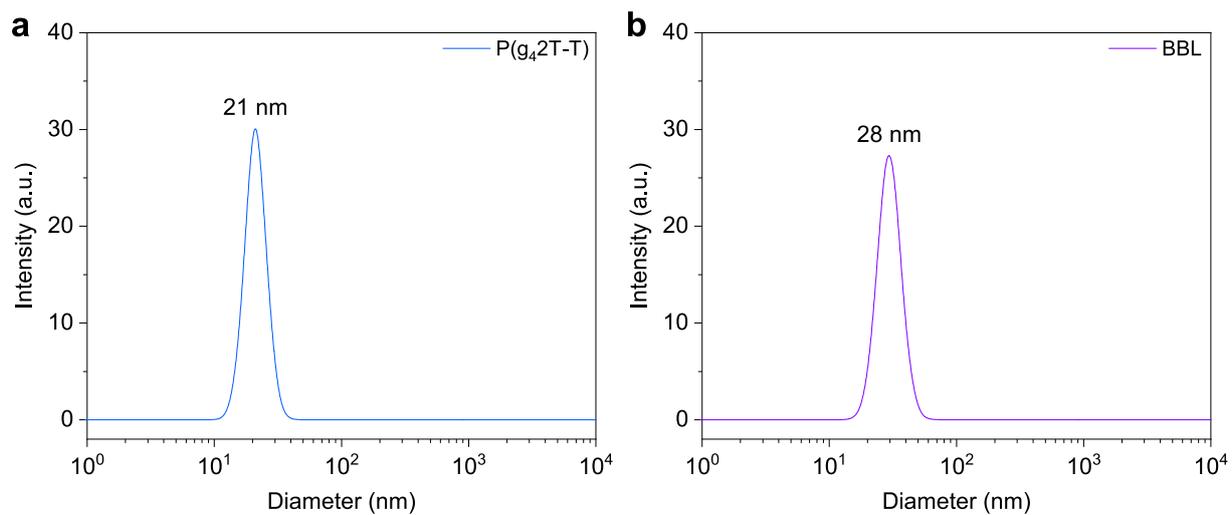

**Supplementary Figure 1 | Ink particle size distribution. a-b**, Dynamic light scattering (DLS) analysis of P(g$_4$2T-T) (**a**) and BBL (**b**) isopropanol-based ink.

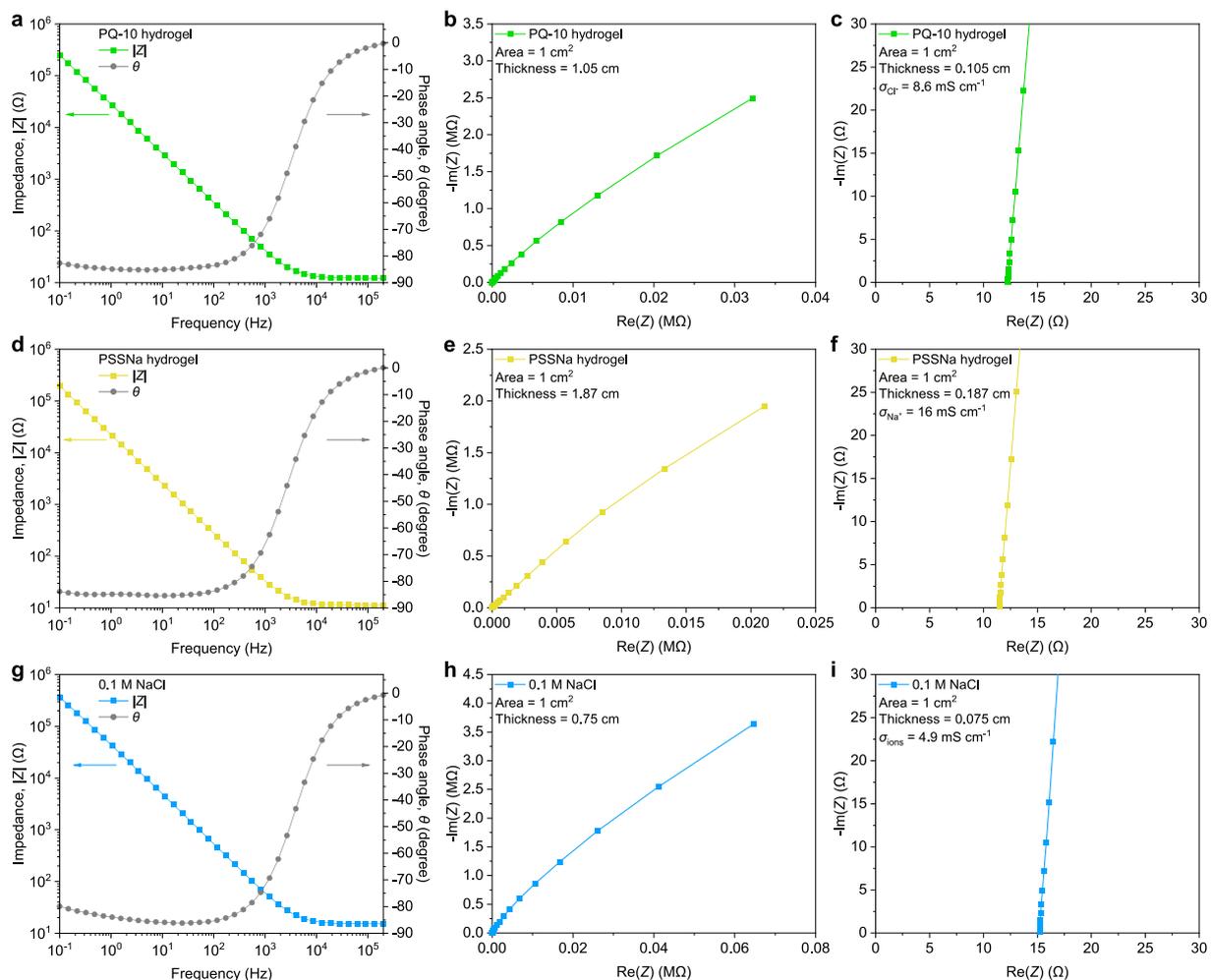

**Supplementary Figure 2 | Ion conductivity. a-c**, Frequency dependent impedance and phase angle (**a**), Nyquist plot (**b**), and Nyquist plot at high-frequency area (**c**) of PQ-10 hydrogel. **d-f**, Frequency dependent impedance and phase angle (**d**), Nyquist plot (**e**), and Nyquist plot at high-frequency area (**f**) of PSSNa hydrogel. **g-i**, Frequency dependent impedance and phase angle (**g**), Nyquist plot (**h**), and Nyquist plot at high-frequency area (**i**) of 0.1 M NaCl aqueous solution. The PQ-10 hydrogel, PSSNa hydrogel, and 0.1 M NaCl have ion conductivity of 8.6, 16 and 4.9 mS cm$^{-1}$, respectively.

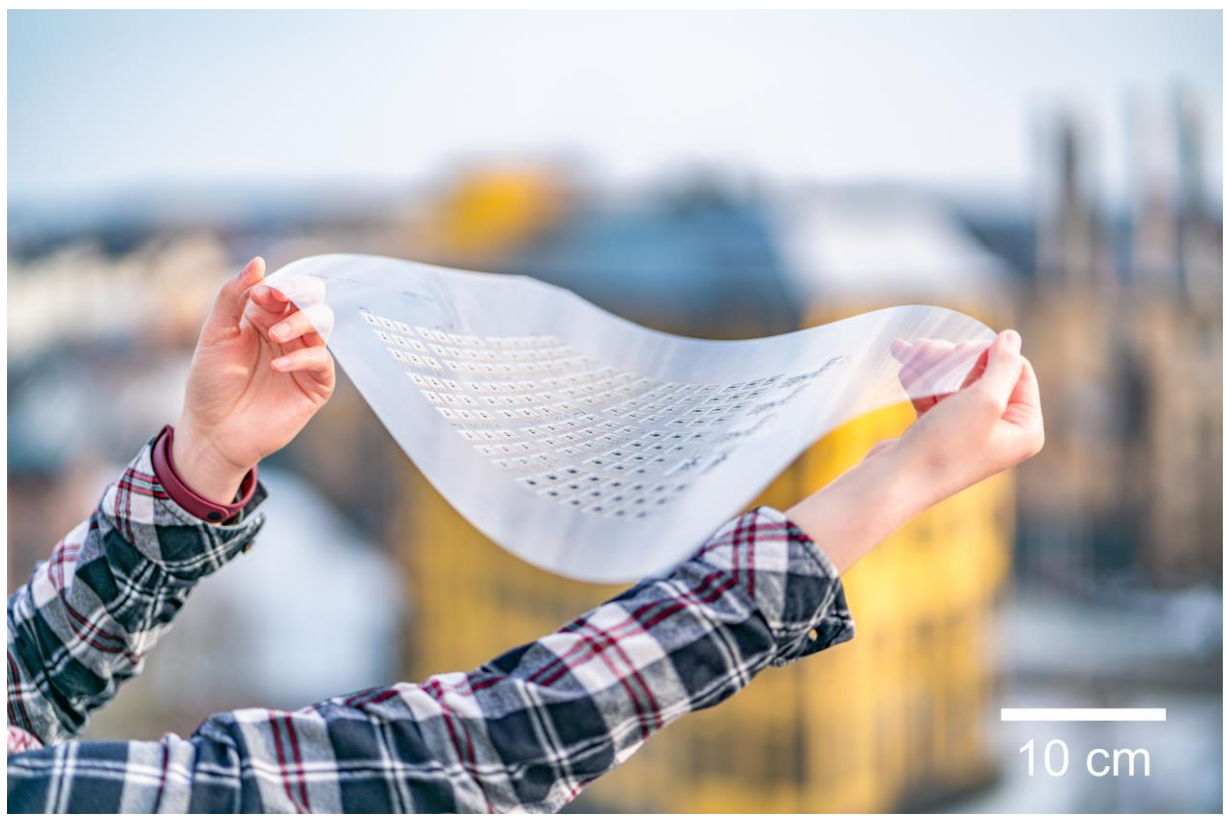

**Supplementary Figure 3 | Printed devices.** Screen-printed carbon/silver electrodes with insulating patterns on an A3-size flexible substrate (PET).

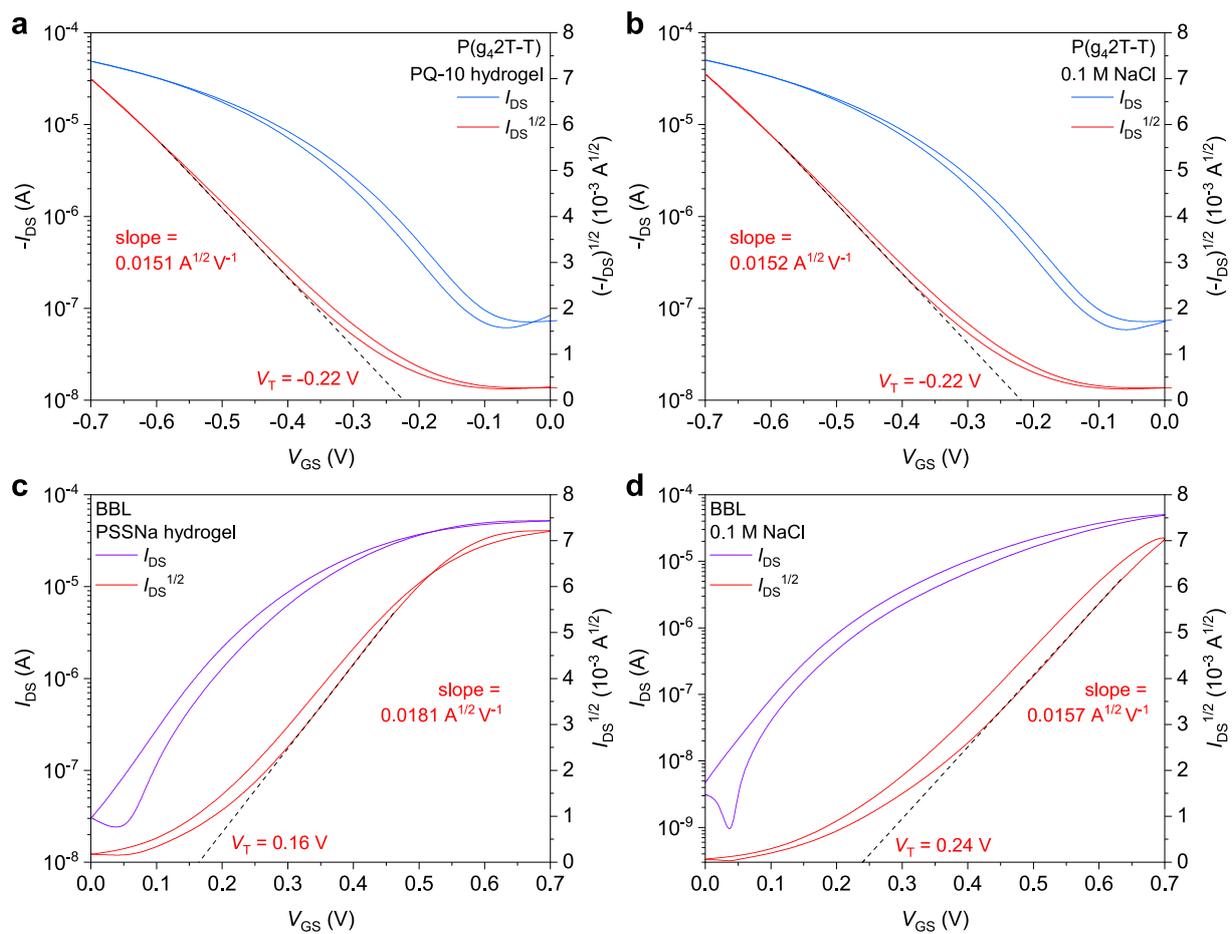

**Supplementary Figure 4 | OECT performance. a-d**, Transfer curve and threshold voltages ($V_T$) of P(g$_4$2T-T) (**a**, **b**) and BBL (**c**, **d**) OECTs, extracted from the transfer characteristics.

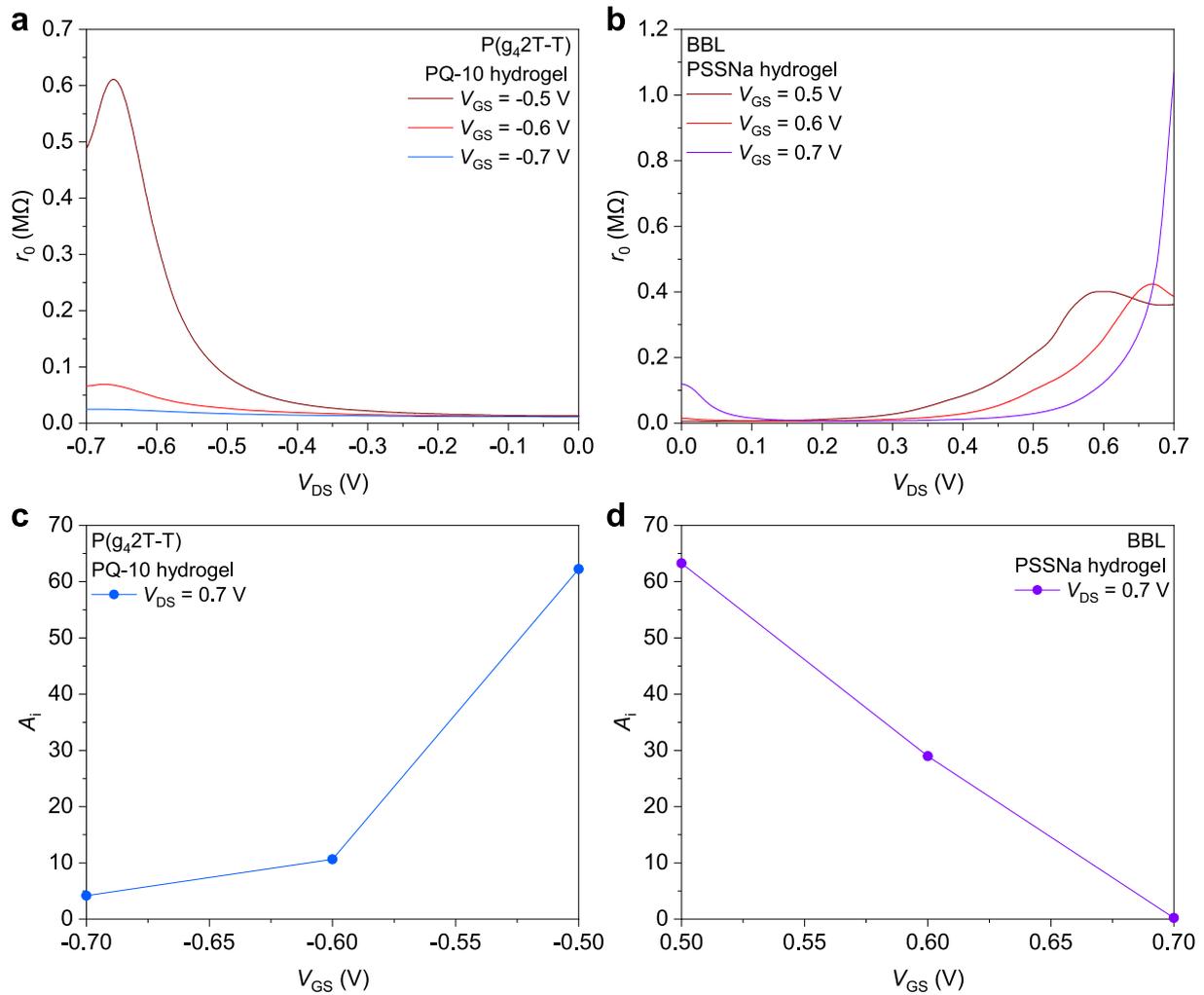

**Supplementary Figure 5 | OECT performance. a-b**, Output resistance ($r_0$) of P(g$_4$2T-T) (**a**) and BBL (**b**) OECTs, derived from the output characteristics. **c-d**, Intrinsic gain of P(g$_4$2T-T) (**c**) and BBL (**d**) OECTs, calculated from product of transconductance ($g_m$) and $r_0$ at the same $V_{DS}$ and $V_{GS}$ level.

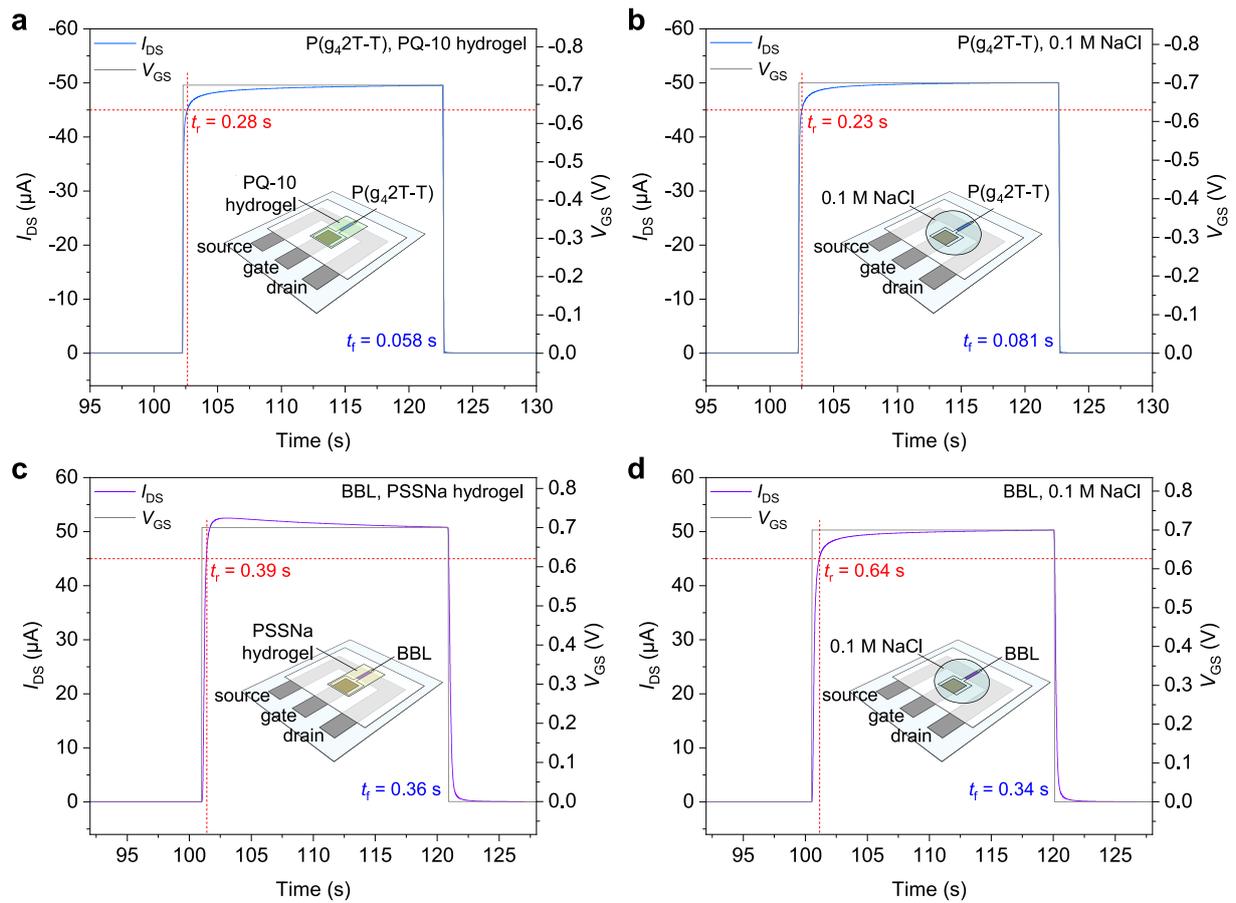

**Supplementary Figure 6 | OECT performance. a-d**, Typical switching characteristics of P(g$_4$2T-T) (**a,b**) and BBL (**c,d**) OECTs, shown rise ($t_r$) and fall ($t_f$) time when applying a voltage pulse at the gate. The definition of rise time is how long it takes to reach 90% of the ON-current while the definition of fall time is how long it takes to fall 90% of the ON-current.

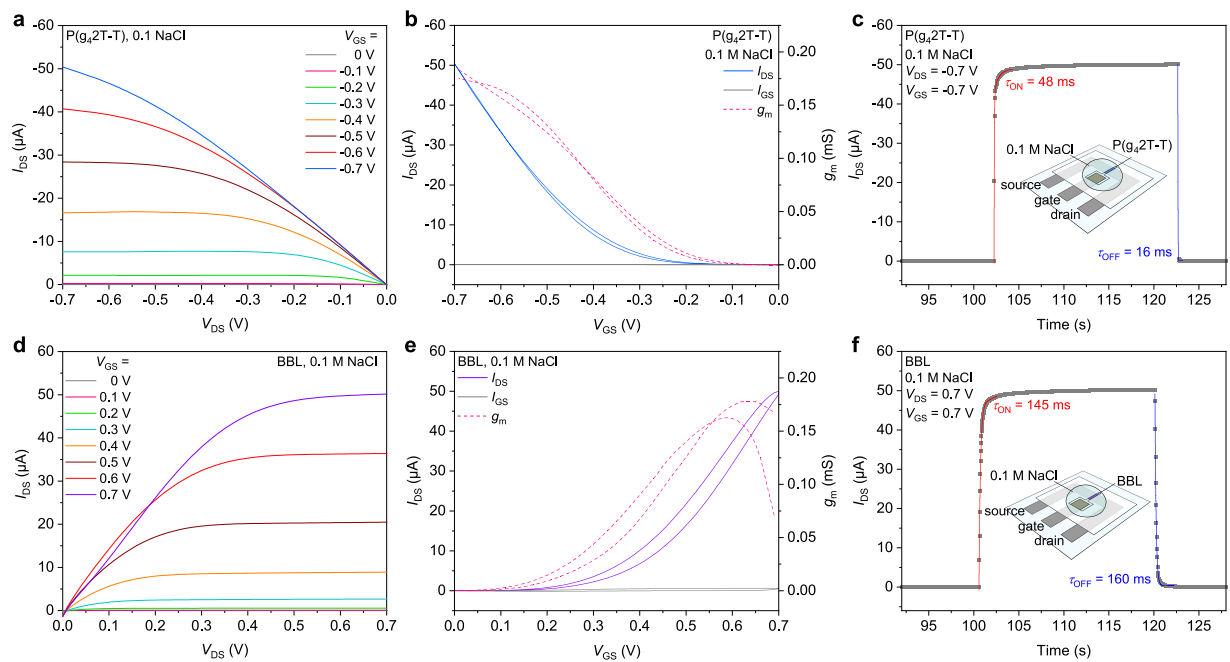

**Supplementary Figure 7 | OECT performance. a-c**, Electrical characteristics of P(g$_4$2T-T) OECT with 0.1 M NaCl as electrolyte: output curve (**a**), transfer curve (**b**), and switching characteristics (**c**). **d-f**, Electrical characteristics of BBL OECT with 0.1 M NaCl as electrolyte: output curve (**d**), transfer curve (**e**), and switching characteristics (**f**).

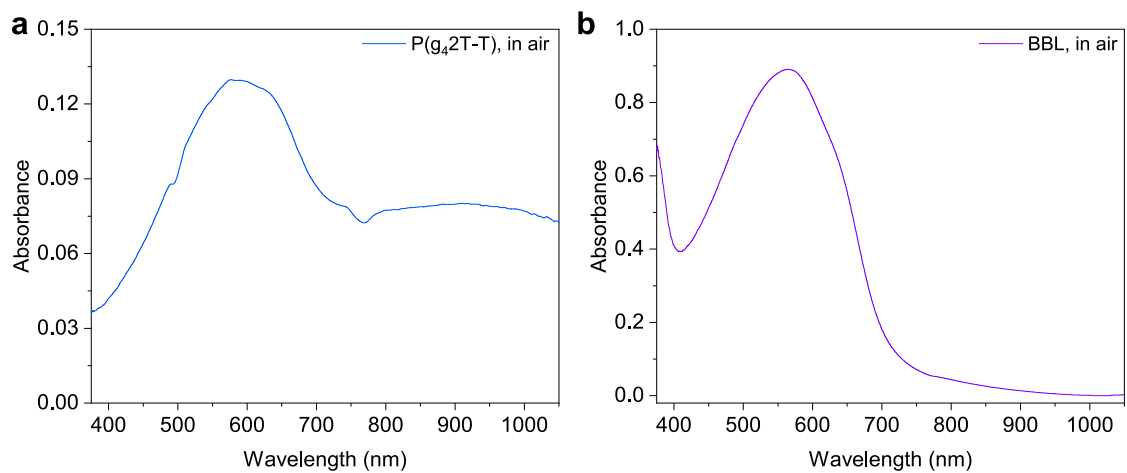

**Supplementary Figure 8 | Absorption spectra. a-b**, Absorption spectra of pristine P(g$_4$2T-T) (**a**) and BBL (**b**) in air.

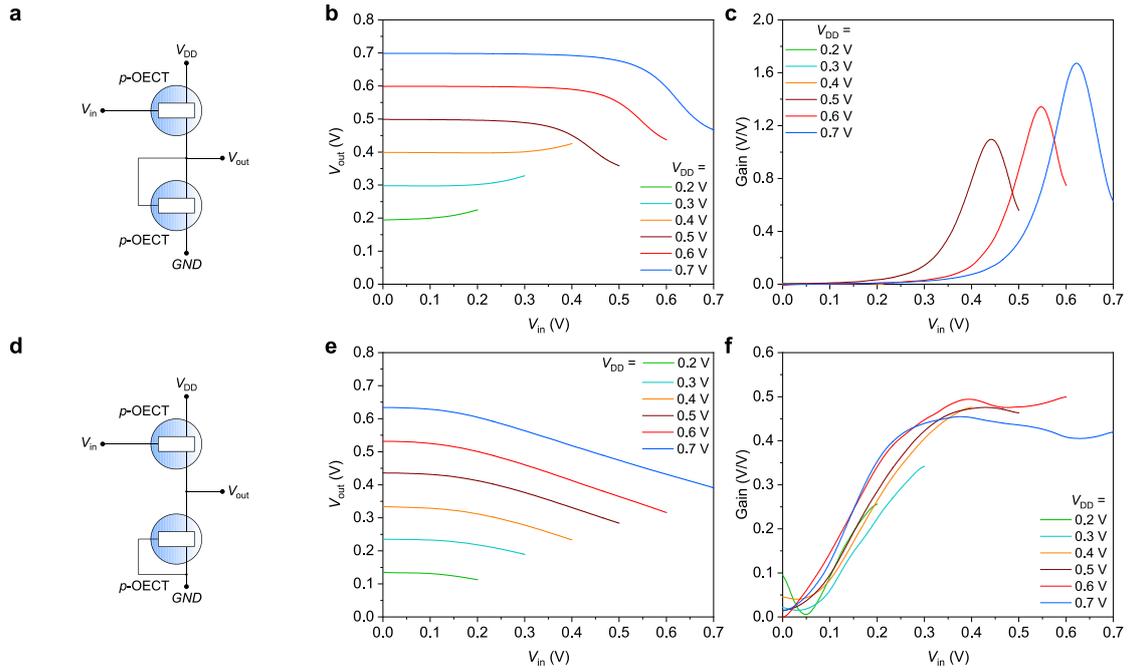

**Supplementary Figure 9 | Inverter performance. a-c**, Electrical characteristics of unipolar inverter based on only p-type P(g$_4$2T-T) OECTs by zero-$V_{GS}$ configuration: device structure diagram (**a**), transfer curve (**b**), and voltage gain (**c**). **d-e**, Electrical characteristics of unipolar inverter based on only p-type P(g$_4$2T-T) OECTs by diode configuration: device structure diagram (**d**), transfer curve (**e**), and voltage gain (**f**).

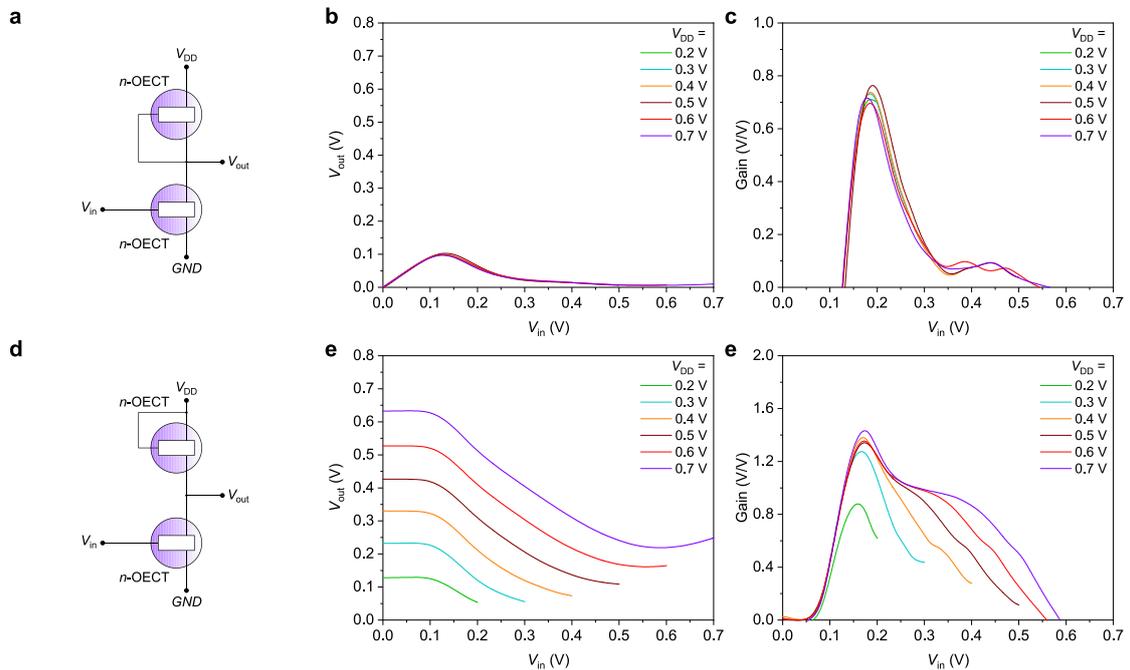

**Supplementary Figure 10 | Inverter performance. a-c**, Electrical characteristics of unipolar inverter based on only n-type BBL OECTs by zero-$V_{GS}$ configuration: device structure schematic diagram (**a**), transfer curve (**b**), and voltage gain (**c**). **d-e**, Electrical characteristics of unipolar inverter based on only n-type BBL OECTs by diode configuration: device structure schematic diagram (**d**), transfer curve (**e**), and voltage gain (**f**).

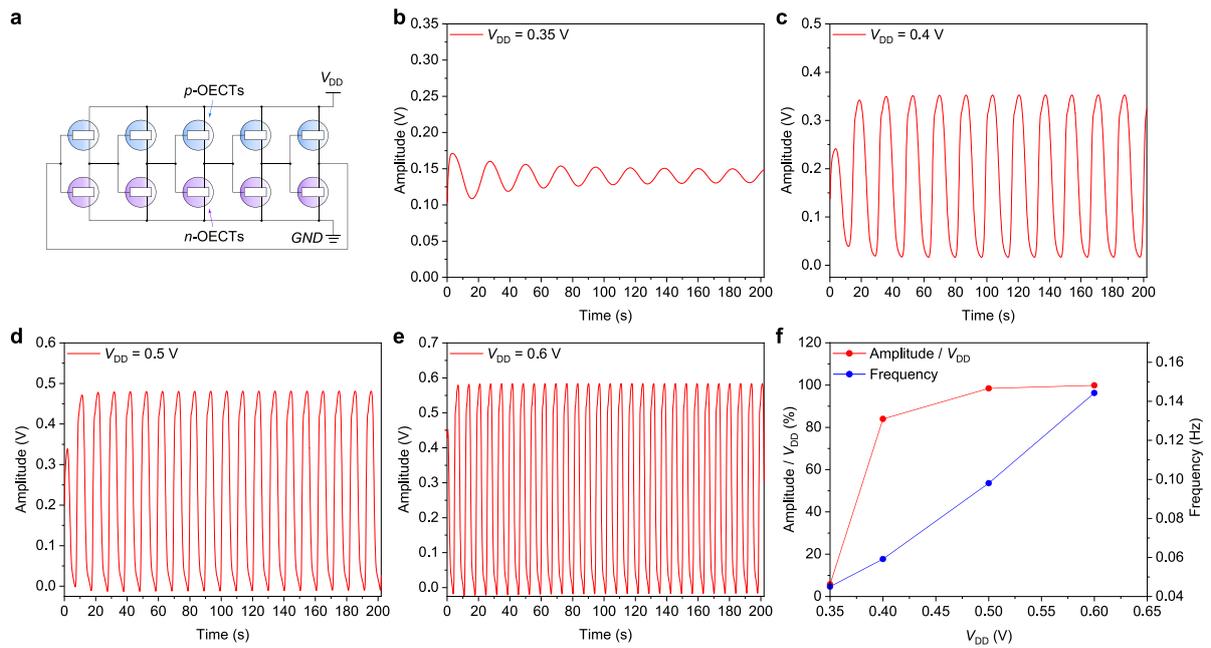

**Supplementary Figure 11 | Ring oscillator performance. a**, Schematic diagram of five-stage ring oscillator. **b-e**, Output voltage waveforms of the 5-stage ring oscillator at $V_{DD}$ = 0.35 V (**b**), 0.4 V (**c**), 0.5 V (**d**), and 0.6V (**e**). **f**, $V_{DD}$ dependent output amplitude and frequency of the ring oscillator.

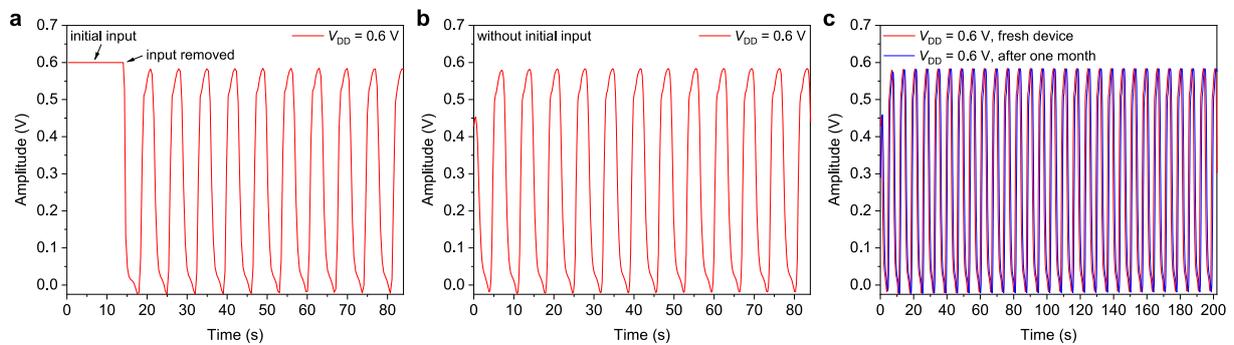

**Supplementary Figure 12 | Ring oscillator performance. a-b**, Output voltage waveforms of the 5-stage ring oscillator with (**a**) or without (**b**) initial input. The ring oscillator can work either with or without initial input. **c**, Stability of the 5-stage ring oscillator. The ring oscillator stored in ambient condition for one month shows consistent performance compared with fresh device.

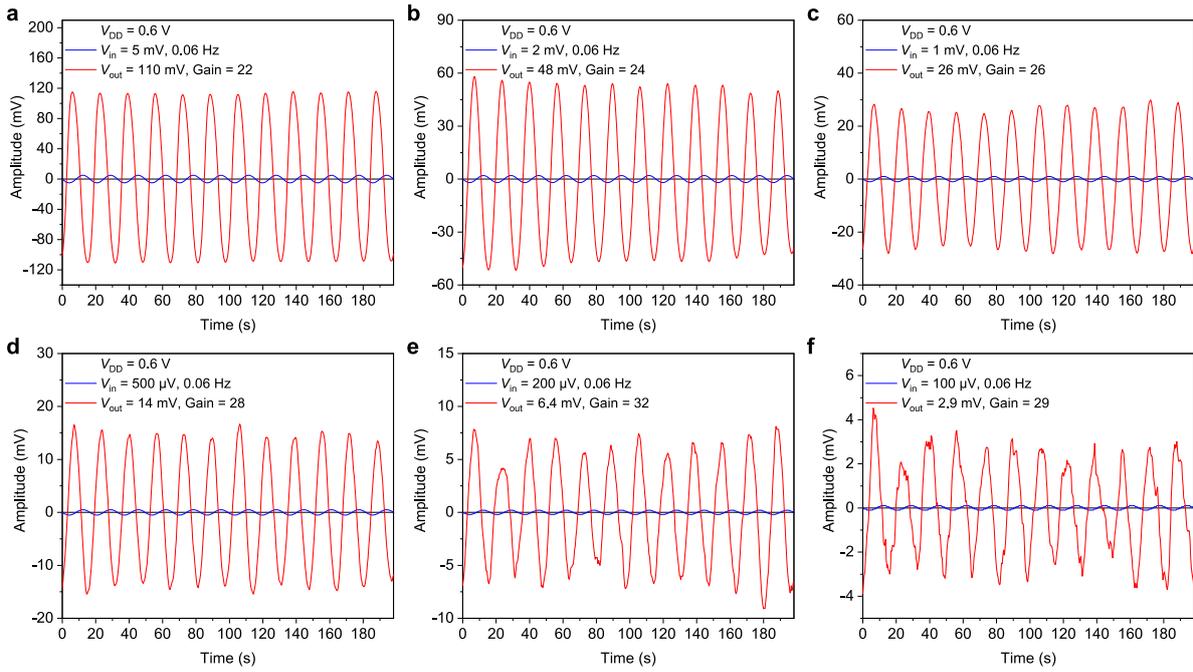

**Supplementary Figure 13 | Voltage amplifier performance. a-f**, Output voltage waveforms of the amplifier with input signals of 0.06 Hz at amplitude = 5 mV (**a**), 2 mV (**b**), 1 mV (**c**), 500 µV (**d**), 200 µV (**e**), and 100 µV (**f**).

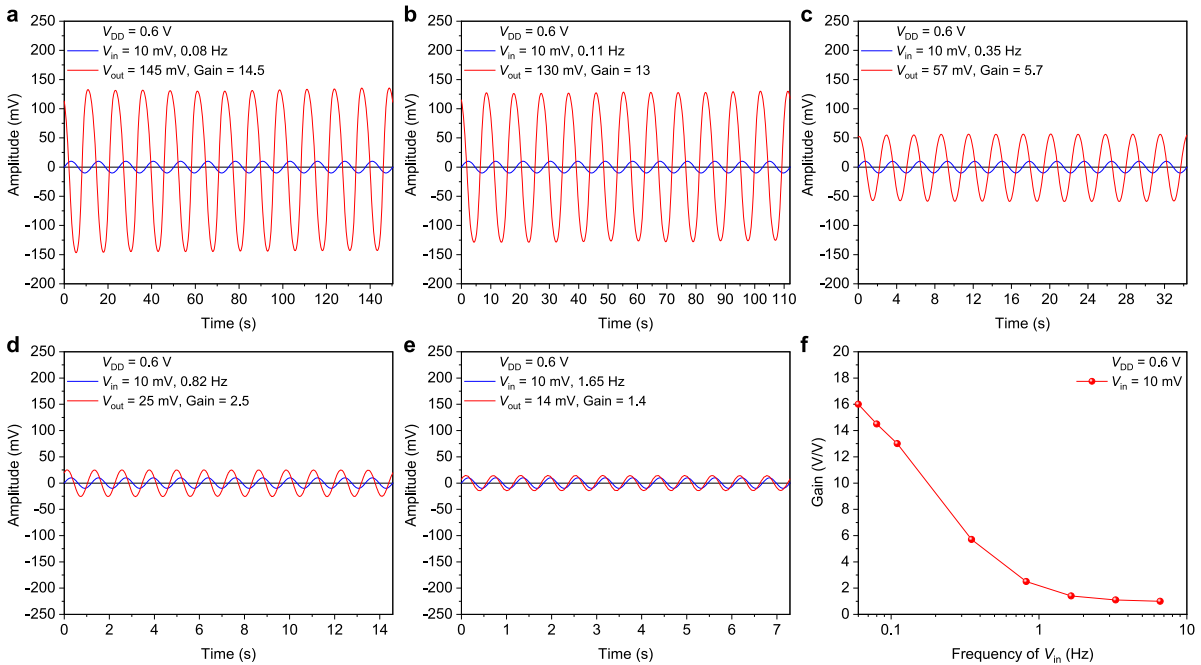

**Supplementary Figure 14 | Voltage amplifier performance. a-e**, Output voltage waveforms of the amplifier with input signals of 10 mV at frequency = 0.08 Hz (**a**), 0.11 Hz (**c**), 0.35 Hz (**d**), 0.82 Hz (**e**), and 1.65 Hz (**f**). **f**, Input frequency dependent voltage gain of the amplifier.

**Supplementary Table 1** Noise margins of the printed complementary OECT inverters at various supply voltages.

| $V_{DD}$ (V) | $V_{IL}$ | $V_{IH}$ | $NM_L$ | $NM_H$ |
|---|---|---|---|---|
| 0.7 | 0.30 | 0.38 | 0.30 | 0.32 |
| 0.6 | 0.24 | 0.32 | 0.24 | 0.28 |
| 0.5 | 0.20 | 0.27 | 0.20 | 0.23 |
| 0.4 | 0.15 | 0.23 | 0.15 | 0.17 |
| 0.3 | 0.11 | 0.18 | 0.11 | 0.12 |
| 0.2 | 0.08 | 0.14 | 0.08 | 0.06 |

**Supplementary Table 2** Comparison of voltage amplifiers among current technologies based on OECTs, electrolyte-gated thin-film transistors, and organic field-effect transistors.

| Technology | Materials | Process | Substrate/ Flexibility | $V_{DD}$ (V) | Gain (V/V, dB) | Power Consumption (µW) | Normalized Gain (dB/µW) |
|---|---|---|---|---|---|---|---|
| *OECTs* | | | | | | | |
| Unipolar | PEDOT:PSS[15] | Photolithography | Glass/No | -6 | 12, 21.6 | 3270 | 0.0066 |
| Unipolar | PEDOT:PSS[16] | Photolithography | Glass/No | -9 | 30, 29.5 | 1350 | 0.022 |
| Unipolar | P(g2T-TT)[17] | Photolithography | Glass/No | -2 | 9.5, 19 | 20 | 0.98 |
| Unipolar | Crystallized PEDOT:PSS[18] | Shadow mask $H_2SO_4$ treatment | Quartz/No | 0.8 | 107, 40 | 480 | 0.085 |
| Complementary | P3CPT, BBL[20] | Photolithography | Glass/No | 0.6 | 12, 21.6 | 15.1 | 1.42 |
| Complementary | PEDOT:PSS, BBL[19] | Shadow mask | Glass/No | 0.5 | 38, 31.6 | 10 | 3.16 |
| Complementary | P(g42T-T), BBL[This work] | Screen-printing spray-coating | PET/Yes | 0.7 | 26, 28.3 | 2.7 | 10.2 |
| | | | | 0.6 | 20, 26 | 1.6 | 16.26 |
| | | | | 0.5 | 16, 24 | 0.8 | 30.10 |
| | | | | 0.4 | 12, 21.6 | 0.4 | 53.96 |
| | | | | 0.3 | 7, 16.9 | 0.1 | 169.02 |
| *Electrolyte-gated TFTs* | | | | | | | |
| Unipolar | P3HT[28] | Aerosol-jet printing | PEN/Yes | 1.5 | 7, 16.9 | 38 | 0.44 |
| Unipolar | IGZO[30] | Photolithography | Quartz/No | 0.5 | 3.77, 11.5 | 5 | 2.31 |
| Complementary | P3HT, ZnO[29] | Shadow mask | PET/Yes | 1 | 18, 25.1 | 600 | 0.042 |
| Ambipolar | CNT[31] | Photolithography | $SiO_2$/No | 1 | 20, 26 | 80 | 0.33 |
| *OFETs* | | | | | | | |
| Unipolar | Mesitylene-based (Merck lisicon S1200)[36] | Ink-jet printing | Parylene-C/Yes | 10 | 1.5, 3.5 | 50 | 0.07 |
| Complementary | PTAA, acenebased-diimide[34] | Screen-printing | PEN/Yes | 40 | 30, 29.5 | 40 | 0.74 |
| Unipolar (Differential) | Pentacene[35] | Shadow mask | PEN/Yes | 15 | 5.6, 15 | 105 | 0.14 |
| Unipolar (Differential) | DNTT[32] | Shadow mask | Parylene-C/Yes | 4 | 56, 35 | 7 | 4.99 |
| Unipolar (pseudo-CMOS) | DNTT[33] | Shadow mask | PEN/Yes | 2 | 200, 46 | 10 | 4.60 |

**Supplementary Table 3** Comparison of DC voltage gain among emerging CMOS technologies operated sub-1V, including OECTs, carbon nanotube FETs, graphene FETs, and 2D transition metal dichalcogenide (TMD) FETs.

| | Materials (p-type, n-type) | Process | Substrate/ Flexibility | $V_{DD}$ (V) | DC Gain (V/V) |
|---|---|---|---|---|---|
| OECTs | PEDOT:PSS, BBL[19] | Shadow mask | Glass/No | 0.5 | 38 |
| | P3CPT, BBL[20] | Photolithography | Glass/No | 0.6 | 12 |
| | P(g$_4$2T-T), BBL[This work] | Screen-printing spray-coating | PET/Yes | 0.7 | 193 |
| | | | | 0.6 | 164 |
| | | | | 0.5 | 110 |
| | | | | 0.4 | 21 |
| Carbon Nanotube FETs | SWCNTs, PbS QDs[43] | Photolithography | Glass/No | 0.9 | 76 |
| | SWCNTs[42] | Solution-processed | PI/Yes | 0.8 | 1.8 |
| Graphene FETs | Graphene[44] | Solution-processed | PEN/Yes | 0.9 | 0.9 |
| | | | | 0.6 | 0.8 |
| | | | | 0.3 | 0.5 |
| TMD FETs | WSe$_2$, MoS$_2$[45] | Photolithography | Glass/No | 1 | 34 |
| | MoTe$_2$, MoS$_2$[46] | Photolithography | Glass/No | 1 | 33 |
| | | | | 0.5 | 8 |
| | | | | 0.25 | 3 |
| | MoTe$_2$[47] | Photolithography | Glass/No | 1 | 17 |
| | MoS$_2$[48] | Photolithography | SiO$_2$/No | 0.1 | 2 |
| | WSe$_2$[49] | Photolithography | SiO$_2$/No | 1 | 3 |
| | MoTe$_2$[50] | Photolithography | SiO$_2$/No | 1 | 3 |